\documentclass[nofootinbib,twocolumn,pre,superscriptaddress,10pt,aps,floatfix,longbibliography]{revtex4-2}
\usepackage[english]{babel}
\usepackage{graphicx}
\usepackage{amsmath}
\usepackage{amssymb}
\usepackage[colorlinks,linkcolor=blue,citecolor=blue,urlcolor=blue]{hyperref}
\usepackage{color}
\usepackage{soul}
\usepackage[normalem]{ulem}
\usepackage{xcolor}
\usepackage{makecell}
\usepackage{multirow}
\definecolor{dblue} {RGB}{28,130,185}
\let\oldaddcontentsline\addcontentsline
\newcommand{\stoptocentries}{\renewcommand{\addcontentsline}[3]{}}
\newcommand{\starttocentries}{\let\addcontentsline\oldaddcontentsline}
\definecolor{nred}{RGB}{224,0,0}
\definecolor{nblue}{RGB}{28,130,185}
\definecolor{darkgreen}{rgb}{0,0.60,.2}
\definecolor{pgreen}{rgb}{0,1.0,0.}

\newcommand\redsout{\bgroup\markoverwith{\textcolor{brown}{\rule[0.6ex]{4pt}{1.0pt}}}\ULon}

\setcellgapes{2pt}
\makegapedcells

\newcommand{\vct}[1]{\mathbf{#1}}

\begin{document}
\title{Defects in Wigner crystals: fracton-elasticity duality and vacancy proliferation}

\author{Pawe\l{} Matus}
\email{matus@tauex.tau.ac.il}
\affiliation{School of Physics and Astronomy, Tel Aviv University, Tel Aviv 69978, Israel}
\affiliation{Max Planck Institute for the Physics of Complex Systems, N\"othnitzer Stra{\ss}e 38, 01187 Dresden, Germany}

\begin{abstract}
We develop a low-energy field theory for electrically charged crystals. Using the tools of fracton-elasticity duality, generalized to accommodate the magnetic 1-form symmetry of electromagnetism, we show how the elastic and electromagnetic degrees of freedom couple to the different crystal defects and to one another. The resulting field theory is then used to calculate vacancy-vacancy interaction energy, and to study the consequences of vacancy proliferation. We find that the longitudinal mode, which in a perfect crystal has a finite gap due to plasma oscillations, becomes gapless in the presence of vacancies. Our framework lays a foundation for a study of defect interactions, their collective dynamics, and consequences of defect-mediated melting in charged crystals.
\end{abstract}
\maketitle
\stoptocentries
\section{Introduction} 

Wigner crystal is a strongly correlated phase of matter predicted in the 1930's by Eugene Wigner \cite{Wigner1938}. Working within the paradigmatic ``jellium" model, in which the atomic cores are treated as providing a uniform positively charged background for the electrons, Wigner showed that when the density of electrons falls below a critical value, the strong electrostatic interactions favor a periodic lattice arrangement of the electrons. While early evidence for Wigner crystallization was obtained in systems such as electrons on liquid helium \cite{Grimes_1980} and later in ultra-clean semiconductor heterostructures \cite{Yoon1999,Huang_2014}, it was very recently that striking signatures were observed in transition metal dichalcogenide monolayers and bilayers \cite{Huang_2021,Zhou_2021,Smolenski_2021,Sung2025}, marking a major advance in the field. These findings motivate an inquiry into how the physics of such a phase of matter is modified by the presence of crystal defects, which will inescapably be present in real systems. In this paper, we propose a low-energy effective action for crystals made up of charged particles, show how the different defects (disclinations, dislocations, and vacancies or interstitials) can be taken account of, and analyze how the presence of the defects can affect the long-wavelength physics of charged crystals.

The program outlined above is carried out using the tools of fracton-elasticity duality. Fractons -- emergent particles with nontrivial constraints on their mobility \cite{Nandkishore:2018sel,Pretko:2020cko,Grosvenor:2021hkn,Gromov:2022cxa} -- occupy a special place among the various types of quasiparticle excitations with which condensed matter physics abounds. 
One of the many exotic features of fracton systems is the emergence of unusal ``higher-rank" gauge fields \cite{Pretko:2018jbi,Gromov2019,Pena2023,Jain:2021ibh}, which mediate interactions between the fracton particles. While fractons have been theorized to emerge in a diverse range of platforms, 
one of their best-known realizations is as elastic defects in two-dimensional crystals: there, defects called disclinations are known to be immobile and disclination dipoles (i.e., dislocations) able to move along the dipole direction only. Simultaneously, the aforementioned higher-rank gauge fields have been shown to arise as duals of the elastic degrees of freedom \cite{Kleinert1982,Kleinert1983,Zaanen_2004,Beekman_2017a,Pretko2018a,Pretko:2018jbi,Pretko2019,Caddeo_2022,Surowka2024a,Surowka2024b}. 

Following the pioneering work on two-dimensional crystals, different versions of the fracton-elasticity duality have also been proposed for three-dimensional crystals \cite{Beekman_2017b,Pretko2018b}, Cosserat (also known as micropolar) crystals \cite{Surowka2020,Hirono:2021lmd}, quasicrystals \cite{Surowka2021,Gaa_2021}, supersolids \cite{Pretko2018c,Kumar2019}, and vortex lattices \cite{Nguyen_2020,Nguyen_2024b,Radzihovsky_2024}. The present work extends this list by deriving the duality for crystals composed of electrically charged particles.
Although some generalisations of the fracton-elasticity duality to charged crystals are present in the literature \cite{Beekman_2017a,Beekman_2017b}, these works suffer a serious limitation: while they take into account some defects, namely dislocations and disclinations, they entirely neglect vacancies and interstitials. The reason for this omission is that while disclinations and dislocations are topological defects, vacancies/interstitials in regular (uncharged) crystals are not, making them difficult to define and treat analytically \cite{Mermin1979,Cvetkovic_2006}. At the same time, however, these defects might be the most important ones to take into account. Indeed, while the energy cost of a dislocation or a disclination grows with the system size, the energy needed to create a vacancy or an interstitial is finite \cite{Fisher1979,Cockayne1990,Candido2001}, meaning that a finite density of them will be present at any non-zero temperature. Moreover, some works \cite{Barraza1999,Pankov2008,Kivelson2024} suggest that even at zero temperature, the two-dimensional Wigner crystal can undergo self-doping such that a finite density of interstitials is present in the ground state. 

Another motivation for studying proliferation of charged defects comes from a recent theoretical work \cite{glodkowski2025} which studies condensation of vacancies/interstitials in a certain class of two-dimensional materials dubbed ``incompressible crystals". The analysis in \cite{glodkowski2025} shows that condensation of vacancies/interstitials in the incompressible crystal leads to the emergence of a gapless longitudinal sound mode. Furthermore, it was suggested that the Wigner crystal might belong to the incompressible crystal class. However, vacancies/interstitials in a charged crystal interact with the electromagnetic gauge field, so their condensation can na\"ively be expected to gap out one gauge field via the Higgs mechanism rather than to add a gapless mode. Thus, it is desirable to find whether the above-described ``anti-Higgs mechanism" does take place in Wigner crystals, and if so, to explain it on physical grounds.

In this work, we study a simple jellium model of a charged crystal, in which the electron spin is neglected. It is known that below a critical density, electrons crystallize into a triangular lattice in two dimensions and into a body-centered cubic lattice in three dimensions \cite{Wigner1938,Fuchs1935,Bonsall1977,Ceperley1980,Ceperley1989}. Here, we focus on the three-dimensional case for a somewhat technical reason: the fact that both the particles and the electromagnetic field inhabit the same number of dimensions makes the analysis more elegant and straightforward. Nevertheless, the duality can be expected to hold with appropriate modifications also in the situation, in which the lattice is two-dimensional, but the electromagnetic field extends into the third dimension, as is the case in the recent experimental realizations of Wigner crystals.

Our derivation of the dual action is based on identifying symmetries of a Wigner crystal, with a special role played by the magnetic 1-form symmetry \cite{Gaiotto_2015,Gomes_2023}. This approach allows us to identify vacancies/interstitials as topologically nontrivial configurations of the dual electromagnetic $U(1)$ gauge field, setting them on an equal footing with the other topological defects: dislocations and disclinations\footnote{Another system where vacancies/interstitials are topological is a superfluid vortex lattice, which resembles a Wigner crystal because of boson-vortex duality \cite{Popov_1973,Nguyen_2020,Nguyen_2024b,Radzihovsky_2024}.}. The derivation of the appropriate effective action and its dual is the subject of Sections~\ref{sec:symmetries}-\ref{sec:dual}. In Sec.~\ref{sec:screening}, the dual theory is put to use in order to find the electric field and the stress field sourced by vacancies/interstitials. Along the way, we find the intraspecies long-range interaction to be repulsive, with the interaction energy varying as $r^{-3}$. Finally, in Sec.~\ref{sec:higgs}, we study the consequences of hypothetical vacancy/interstitial condensation, demonstrating the anti-Higgs mechanism, and explaining its origin. It is found to be closely linked to the phenomenon of plasma oscillations, which create a gap for the longitudinal modes in the absence of defects, but in the presence of defects can be rendered inactive.

The paper makes use of the following conventions. The electron charge is set to $-1$. Greek indices denote both temporal and spatial directions, e.g., $A_\mu \in \{A_t,A_x,A_y,A_z\}$, Latin indices denote spatial directions, e.g., $A_i \in \{A_x,A_y,A_z\}$, and spatial vectors are occasionally denoted by bold letters, e.g., $\vct E = (E_x,E_y,E_z)$. The bold $\vct x$ refers to the position in space: $\vct x = (x,y,z)$, while the non-bold $x$ refers to the position in both time and space: $x=(t,x,y,z)$. We assume throughout that the metric is Euclidean, meaning that no distinction is made between upper and lower indices, and we employ the Einstein summation convention for repeating indices, e.g., $A_iB_i = A_xB_x+A_yB_y+A_zB_z$. The Hodge dual $*$ is also taken with respect to the Euclidean metric. Finally, index symmetrisation is denoted by round brackets: $A_{(ij)}=\frac{1}{2}\left(A_{ij}+A_{ji}\right)$, antisymmetrisation by square brackets: $A_{[ij]}=\frac{1}{2}\left(A_{ij}-A_{ji}\right)$, and taking the symmetric traceless part by angle brackets: $A_{\langle ij \rangle}=A_{(ij)}-\frac{1}{3}\delta_{ij}A_{kk}$.

\section{Symmetries of a Wigner crystal}
\label{sec:symmetries}

\begin{figure}
    \includegraphics[width=0.8\linewidth]{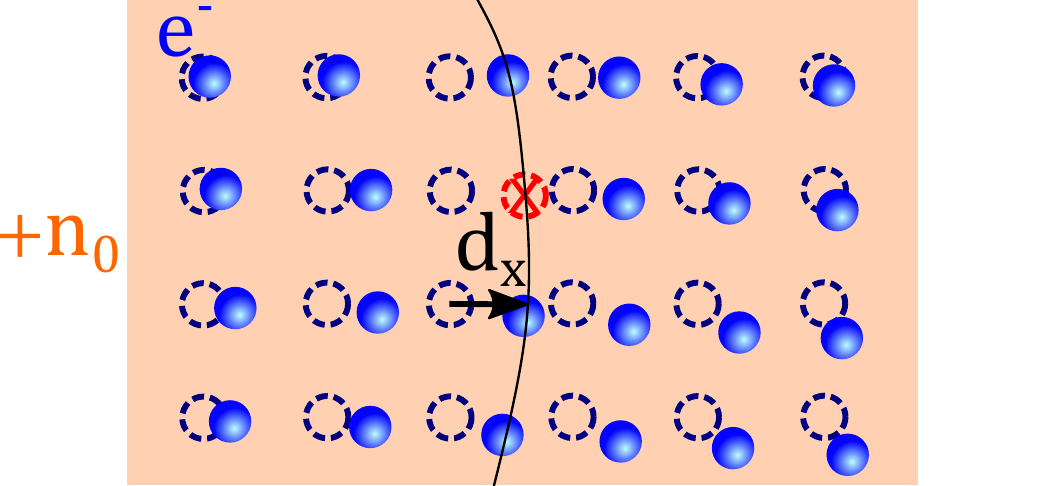}
    \caption{A cross-section view of a Wigner crystal. The electrons $e^-$ form a lattice inside a background medium with a uniform charge density $+n_0$. The displacement of the electrons (full blue spheres) from their equilibrium positions (dashed blue circles) is given by $d_i=x_i-X_i$, with $d_x$ along one vertical line visualized in the plot. The dashed red circle with the $X$ mark represents a vacancy.}
    \label{fig:deformations}
\end{figure}

We adopt a simplified view of a Wigner crystal as a lattice of charged particles residing in a uniform, neutralizing background with charge density $n_0$ -- see Fig.~\ref{fig:deformations}. In order to develop the low-energy, coarse-grained description of such a system, we first identify the appropriate degrees of freedom. 

As in generic crystals, one set of fundamental fields is given by the three comoving (Lagrangian) coordinates $X_i(x)$, which assign to the solid element found at $x$ the position it occupies in an equilibrium \cite{Son2005,Nicolis:2013lma}. Thus, the displacement field is given by $d_i(x) = x_i-X_i(x)$.

The electrons interact with the dynamical electromagnetic field, which introduces another set of fundamental fields. Here, we choose to express the electric field $\vct E$ and the magnetic field $\vct B$ through the ``dual" gauge field one-form $\kappa = \kappa_t dt + \kappa_i dx_i$ such that
\begin{equation}
    \vct E = \nabla \times  \boldsymbol \kappa,~~~~\vct B = -\nabla \kappa_t+\partial_t\boldsymbol\kappa.
\end{equation}
If $\kappa$ is single-valued, the identities $\nabla \vct E=0$ and $\partial_t\vct E-\nabla\times\vct B=0$ preclude the presence of electric charges and currents. Electric charges can, however, be captured by topologically nontrivial $\kappa$ in the same way that magnetic monopoles can be captured by topologically nontrivial ``standard" gauge fields. 

In the presence of the electromagnetic field, the action possesses the magnetic 1-form symmetry $U(1)_m^{(1)}$\footnote{The electric one-form symmetry $U(1)_e^{(1)}$ is explicitly broken by the presence of dynamical charged matter.}, which is naturally expressed in terms of the dual gauge field $\kappa$. The charged objects of the theory are 't Hooft loops $T[\mathcal{C}]$ supported along one-dimensional closed lines in spacetime $\mathcal{C}$ and defined as
\begin{equation}
    T[\mathcal{C}]=\exp\left(i\int_\mathcal{C} \kappa\right).
\end{equation}
The $U(1)_m^{(1)}$ transformation is parametrized by a one-form $\xi$ such that $d\xi=0$. Denoting the magnetic one-form symmetry generator by $\mathcal{Q}^{(1)}$, the symmetry can be expressed as \cite{Gomes_2023}
\begin{equation}
    \exp\left(\xi \mathcal{Q}^{(1)}\right): T[\mathcal{C}]\rightarrow \exp\left(i\int_\mathcal{C} \xi\right)T[\mathcal{C}].
    \label{eq:magnetic_symmetry}
\end{equation}

\begin{figure}
    \includegraphics[width=0.45\linewidth]{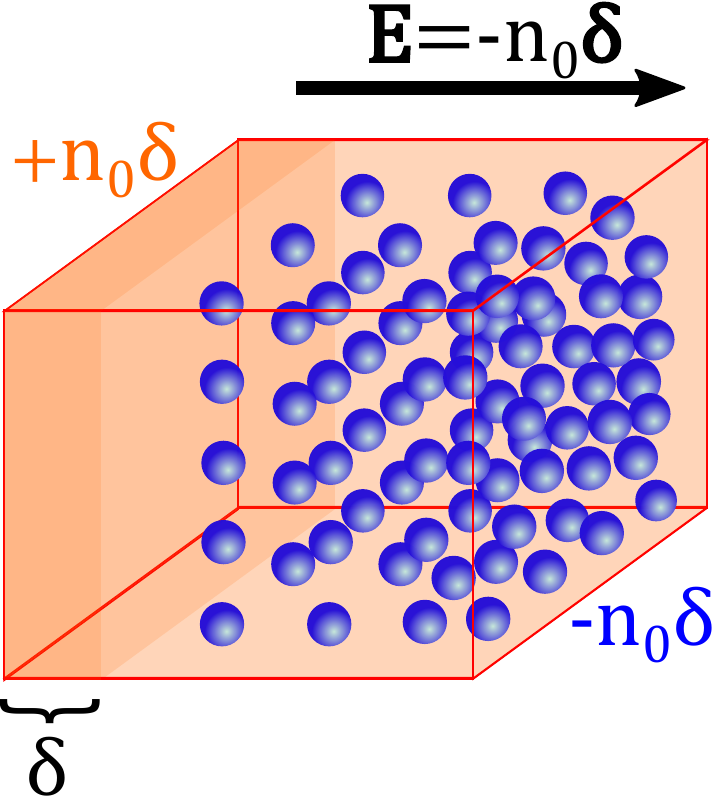}
    \caption{A translation of the charged background relative to the electrons by $\boldsymbol\delta$ creates surface charge $\pm n_0\delta$ on the opposite sides, which in turn induces the electric field $\vct E = -n_0\boldsymbol\delta$.}
    \label{fig:displacement}
\end{figure}

Consider now a ``lattice translation'' by a constant vector $\boldsymbol \delta$: $X_i(x)\rightarrow X_i(x)+\delta_i$. This transformation can be understood as moving the background relative to the electrons by a vector $\boldsymbol\delta$. While in neutral crystals such a shift is inconsequential, in Wigner solids it produces a uniform electric field $\vct E = -n_0\boldsymbol\delta$ because of the charge accumulating on the surface: see Fig.~\ref{fig:displacement}. Importantly, this phenomenon survives taking the thermodynamic limit, when the surfaces are removed to infinity. Let us then define the modified lattice translation $T_i$ such that the action of $\exp(\delta_iT_i)$ is
\begin{equation}
\exp(\delta_iT_i):
    \begin{cases}
        X_i(x) &\rightarrow X_i(x)+\delta_i, \\
        \boldsymbol\kappa(x)&\rightarrow \boldsymbol\kappa(x) + \frac{n_0}{2}\boldsymbol\delta\times\vct x,\\
        \kappa_t(x)&\rightarrow \kappa_t(x).
    \end{cases}
    \label{eq:Ti_symmetry}
\end{equation}
This can be understood as simultaneously shifting the charged background and applying an additional electric field $\vct E=n_0\boldsymbol{\delta}$ to compensate for the electric field created by this shift. Following the above discussion, we postulate our action to be symmetric under $T_i$. 

While lattice translations take on a rather unusual shape in Wigner crystals, the other crystal symmetries are more generic. One of them is the ``space translation" symmetry generated by $P_i$:
\begin{equation}
\exp(\delta_iP_i):
    \begin{cases}
        X_i(t,\vct x) &\rightarrow X_i(t,\vct x-\boldsymbol\delta)+\delta_i, \\
        \kappa(t,\vct x)&\rightarrow \kappa(t,\vct x-\boldsymbol\delta).
    \end{cases}
    \label{eq:Pi_symmetry}
\end{equation}
This corresponds to shifting both the electrons and the background by the same vector $\boldsymbol\delta$. Then, there is the time translation symmetry $H$:
\begin{equation}
\exp(\delta_t H):
    \begin{cases}
        X_i(t,\vct x) &\rightarrow X_i(t-\delta_t,\vct x), \\
        \kappa(t,\vct x)&\rightarrow \kappa(t-\delta_t,\vct x).
    \end{cases}
    \label{eq:H_symmetry}
\end{equation} 
Finally, there are two distinct rotation symmetries: the ``lattice rotation'' generated by $L_i=\epsilon_{ijk}X_j T_k$, and the ``space rotation'' generated by  $J_i=\epsilon_{ijk}x_jP_k$. 

\section{Effective action}
\label{sec:effective}

The ground state of the system in the crystalline phase breaks several of the symmetries discussed in Sec.~\ref{sec:symmetries}. Each of the broken continuous symmetries supplies a Goldstone-like field (called a Stueckelberg field), in terms of which the low-energy dynamics of the system can be expressed. Thus, we begin by identifying the broken symmetries and the corresponding low-energy fields. 

In equilibrium the Lagrangian coordinates are equal to the Eulerian coordinates, $X_i(x)=x_i$. This configuration breaks $T_i$ and $L_i$, but leaves $P_i$ and $J_i$ unbroken. In addition, $\mathcal{Q}^{(1)}$ is broken in the Coulomb phase. Thus, the symmetry breaking pattern is
\begin{equation}
    \begin{split}
        \mathrm{Unbroken:}&\qquad H, P_i, J_i, \\
        \mathrm{Broken:}&\qquad T_i, L_i, \mathcal{Q}^{(1)}.
    \end{split}
\end{equation}
With this insight, we obtain the fundamental degrees of freedom as follows.
\begin{itemize}
    \item We take the negative displacement field $u_i(x)=-d_i(x)=X_i(x)-x_i$ as the field corresponding to the broken $T_i$.
    \item Let us denote by $\kappa^{\mathrm{m}}(\vct x)$ the multi-valued Dirac vector potential field of a monopole, which satisfies $d^2 \kappa^{\mathrm{m}}(\vct x)=\delta^{(3)}(\vct x)$. Out of equilibrium, the lattice sites gives rise to the singular vector potential 
    \begin{equation}
        \kappa^{\mathrm{latt}}(x) =  -\sum_{j=1}^N\left[\kappa^{\mathrm{m}}(\vct x-\vct x_j(t))-\kappa^{\mathrm{m}}(\vct x-\vct X_j)\right],
    \end{equation}
    where $\vct x_j(t)$ is the position of site $j$ at time $t$ and $N$ is the total number of sites. We then split the full gauge field $\kappa$ into $\kappa^{\mathrm{latt}}$
    and the remaining part $\phi$, which just like $\kappa$ is a one-form with one temporal and three spatial components $\phi = \phi_t dt + \phi_i dx_i$\footnote{A similar splitting was performed for a vortex lattice in superfluids in Ref.~\cite{Watanabe_2013}.}:
    \begin{equation}
        \phi(x) = \kappa(x)-\kappa^{\mathrm{latt}}(x).
    \end{equation}
    With this splitting, $\kappa^{\mathrm{latt}}(x)$ is determined solely by the positions of the lattice sites, meaning that its dynamics is tied to the dynamics of the lattice, while $\phi(x)$ is an independent degree of freedom corresponding to the broken $\mathcal{Q}^{(1)}$. Note that in the presence of charged defects, not captured by the field $X_i(x)$, $\phi(x)$ will also become singular.
    \item Since the translational symmetry is broken, there is no need to specify a Stueckelberg field associated with the broken rotational symmetry, since such a field is not independent from $u_i$, see \cite{Low2002,Watanabe_2013}. This is also known from the physics of regular crystals, where the number of phonon modes is equal to the number of crystal dimensions\footnote{ Including a field quantifying local rotations as a separate degree of freedom can, however, be used to construct the effective action for Cosserat elasticity \cite{Surowka2020}.}. 
\end{itemize}
The building blocks of the effective action are the combinations of $u_i$ and $\phi_\mu$ that are invariant under $T_i$ and $\mathcal{Q}^{(1)}$. Such combinations are
\begin{equation}
     2\partial_{[t} \phi_{i]},\quad2\partial_{[i}\phi_{j]}-n_0\epsilon_{ijk}u_k,\quad \partial_tu_i,\quad\partial_j u_i.
     \label{eq:invariants}
\end{equation}
While we have introduced the low-energy fields and their invariant combinations in a somewhat ad hoc way, in App.~\ref{app:coset} we find the same objects via a more rigorous procedure known as the coset construction \cite{Volkov1973,Ogievetsky1974,Low2002,Landry_2020,Hirono:2021lmd}.

Taking into account also the rotational symmetries, the most general quadratic action can be written as
\begin{equation}
\begin{split}
    S = & \frac12 
    \int d^4 x \left[c_B (2\partial_{[t}\phi_{i]})^2 +c_{u}\left(\partial_t u_i\right)^2\right.\\& 
    \left.- \frac{c_E}{2}\left(2\partial_{[i}\phi_{j]}-n_0\epsilon_{ijk}u_k\right)^2-C_{ijkl}\partial_i u_{j}\partial_ku_{l}\right],
    \label{eq:action}
\end{split}
\end{equation}
where $c_B$, $c_E$, $c_u$ are some positive coefficients and $C_{ijkl}$ is a rank-four tensor. Physically, we can identify $c_u$ with the mass density of the electrons, $c_u=m_e n_0$, $C_{ijkl}$ with the elasticity tensor, and the meaning of $c_E$ and $c_B$ shall become clearer in the next section. The most general rank-four tensor consistent with isotropy, parity, and angular momentum conservation takes the form
\begin{equation}
    C_{ijkl} = \frac{C_b}{3} \delta_{ij}\delta_{kl}+\frac{C_s}{2}(\delta_{ik}\delta_{jl}+\delta_{il}\delta_{jk}-\frac{2}{3}\delta_{ij}\delta_{kl}),
\end{equation}
where $C_b$ is the bulk elastic modulus and $C_s$ is the shear elastic modulus. The real three-dimensional Wigner crystal is not isotropic, but rather has cubic symmetry, which increases the number of elastic moduli to three, one bulk and two shear \cite{Thomas1966}; nevertheless, for simplicity we assume that the two shear moduli are equal.

In the absence of defects, all the fields are smooth and the equations of motion of the action~(\ref{eq:action}) can be solved exactly. Alternatively, if one is interested in the gapless sector of the theory only, the massive mode $(n_0 u_i-\epsilon_{ijk}\partial_j \phi_k)$ can be first integrated out of the action, which amounts to setting the value of $u_i$ to
\begin{equation}
    u_i \rightarrow n_0^{-1} \epsilon_{ijk}\partial_j \phi_k.
\end{equation}
This leads to an important observation: at low energies, the solid is incompressible, $\partial_iu_i=0$. The long-wavelength action for the gapless sector reads
\begin{multline}
    S_{\mathrm{gapless}} = \frac{1}{2}\int d^4 x\left[c_B (2\partial_{[t}\phi_{i]})^2\right.\\
    \left.-n_0^{-2}C_s\left(\partial_{(i}\epsilon_{j)kl}\partial_{k}\phi_{l}\right)^2\right].
\label{eq:S_gapless}
\end{multline}
We find two gapless transverse modes with the quadratic dispersion relation
\begin{equation}
    \omega = \pm\sqrt{\frac{C_s}{2n_0^2c_B}}k^2.
    \label{eq:dispersion_unbroken}
\end{equation}
This is in stark contrast to regular crystals, in which there are three linear modes, two transverse and one longitudinal. At the same time, our result agrees with previous studies concerning Wigner crystals \cite{Beekman_2017a,Beekman_2017b}. Recently, \cite{Du2024} utilized the magnetic 1-form symmetry to propose a low-energy effective action for a Wigner crystal similar to Eq.~(\ref{eq:S_gapless}).

\section{Dual gauge theory}
\label{sec:dual}

Crystal defects give rise to singularities in the low-energy fields $u_i$ and $\phi=\phi_t dt + \phi_i dx_i$. Thus, in order to study their behavior more closely, we decompose the fields into smooth parts denoted with a bar and singular parts denoted with the superscript $s$:
\begin{equation}
    \phi = \bar\phi+\phi^s, \qquad u_i=\bar{u}_i+u_i^s.
\end{equation}
Both singular fields $\phi^s$ and $u_i^s$ have a clear physical interpretation, which we shall now discuss.

Since the full gauge field $\kappa$ is the dual electromagnetic $U(1)$ gauge field, it exhibits point singularities at the locations of electric monopoles. At the same time, given the splitting $\kappa = \kappa^{\mathrm{latt}}+\phi$ (see Sec.~\ref{sec:effective}), the singularities associated with the lattice sites and the background charge are incorporated into $\kappa^{\mathrm{latt}}$. Consequently, $\phi$ is regular everywhere except at the lattice sites at which an electron is missing, such as the vacancy site in Fig.~\ref{fig:deformations}, or when an excess electron forms an interstitial. To be more precise, if we define the vacancy density as 
\begin{equation}
    \rho^{\mathrm{vac}}\equiv  \partial_i\left(\epsilon_{ijk}\partial_j \phi^s_k\right),
    \label{eq:vacancy_density}
\end{equation}
integrating $\rho^{\mathrm{vac}}$ over a finite volume $V$ gives
\begin{equation}
    \int_V \rho^{\mathrm{vac}} = n^{\mathrm{vac}}-n^{\mathrm{int}},
\end{equation}
where $n^{\mathrm{vac}}$ is the number of vacancies (lattice sites with extra charge $+1$) and $n^{\mathrm{int}}$ is the number of interstitials (excess electrons with charge $-1$) enclosed within $V$. 

\begin{figure}
    \includegraphics[width=0.6\linewidth]{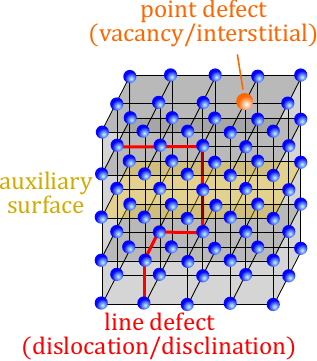}
    \caption{Visualization of a line defect (red line) and a point defect (orange sphere). An integral of the line defect density over the auxiliary surface (yellow) yields a topological invariant.}
    \label{fig:dislocation}
\end{figure}

Let us now turn our attention to the singular part of the displacement field $u_i$. Displacements differing by a linear combination of lattice vectors describe the same lattice, meaning that we can identify $u_i\sim u_i+a_i$, where $a_i$ is a lattice vector. Thus, the order parameter manifold is a 3-torus, and $u_i$ has line singularities classified by an element of $\mathbb{Z}^3$. We define the dislocation density as the $i$-indexed vector $\boldsymbol{\rho}^{\mathrm{disl}}_i = (\rho^{\mathrm{disl}}_{ix},\rho^{\mathrm{disl}}_{iy},\rho^{\mathrm{disl}}_{iz})$:
\begin{equation}
    \boldsymbol{\rho}^{\mathrm{disl}}_i\equiv \nabla\times\left(\nabla u^s_i\right).
    \label{eq:dislocation_density}
\end{equation}
Let the vector $\boldsymbol{\rho}^{\mathrm{disl}}_{i}$ intersect a bounded surface $S$ characterized by the normal vector $\hat{\vct n}$, see Fig.~\ref{fig:dislocation}. The surface integral gives
\begin{equation}
    \int_S \boldsymbol{\rho}^{\mathrm{disl}}_{i}\cdot \hat{\vct n} =  b_i,
    \label{eq:burgers}
\end{equation}
where the Burgers vector  $b_i$ is defined as the total change in the displacement field $u_i$ as one travels around the boundary of the surface $S$. For the displacement field to be well-defined, the Burgers vector must be a linear combination of lattice vectors.

In addition to interstitials/vacancies and dislocations, there exists one more type of topological defects known as disclinations, which represent singularities in the ``bond angle vector" $\Theta_i$. 
While a proper characterization of such defects would have to take into account the non-Abelian nature of the rotation group, here we will assume, as is customary \cite{Kleinert1989}, the description appropriate for continuous isotropic media with infinitesimal Frank vectors. In that case, the ``bond angle vector" is defined as
\begin{equation}
    \Theta_i \equiv \frac{1}{2}\epsilon_{ijk}\partial_j u_k,
    \label{eq:bond_angle}
\end{equation}
and similarly to dislocations, the disclination density is the $i$-indexed vector
\begin{equation}
    \boldsymbol{\rho}^{\mathrm{disc}}_i\equiv \nabla\times\left(\nabla \Theta^s_i\right).
    \label{eq:disclination_density}
\end{equation}
The Frank vector can be obtained by integrating $\boldsymbol{\rho}^{\mathrm{disc}}_i$ over a surface, similar to Eq.~(\ref{eq:burgers}).
Since both disclinations and dislocations are defined via the singular component of the displacement $u_i^s$, they are closely related: a disclination can be seen as a termination of an infinite row of dislocations, while a dislocation can be seen as a bound state of two disclinations with opposite Frank vectors \cite{Kleinert1989}. Relatedly, an isolated disclination line adds a finite-density contribution to the dislocation density $\boldsymbol{\rho}^{\mathrm{disl}}_i$ defined in Eq.~(\ref{eq:dislocation_density}). 

As we shall see, the defect densities~(\ref{eq:vacancy_density}),~(\ref{eq:dislocation_density}) and ~(\ref{eq:disclination_density}) emerge in a natural way when the methods of gauge-elasticity duality \cite{Kleinert1982,Pretko2019,Surowka2020,Hirono:2021lmd} are applied to the elastic action~(\ref{eq:action}). The dual language offers a new perspective on the theory of elasticity: the singular field configurations become currents coupled to gauge fields, whose dynamics are related to those of the smooth fields $\bar{\phi}$ and $\bar{u}_i$. As described in more detail in App.~\ref{app:duality}, two gauge fields emerge by dualizing $\bar{\phi}$ and $\bar{u}_i$: a 1-form field $A^{(\phi)}=A^{(\phi)}_\mu dx_\mu$ and a vector-valued 2-form field $A^{(u)}_i=\frac12A^{(u)}_{i\mu\nu} dx_\mu\wedge dx_\nu$ that transform under the gauge symmetry
\begin{equation}
\begin{split}
    A^{(\phi)}&\rightarrow A^{(\phi)}+d\lambda^{(\phi)},\\
    A^{(u)}_i&\rightarrow A^{(u)}_i+d\lambda^{(u)}_i+\frac12n_0\epsilon_{ijk}dx_j\wedge dx_k \lambda^{(\phi)},
\end{split} \label{eq:gauge}
\end{equation}
for a 0-form (scalar function) $\lambda^{(\phi)}$ and a 1-form $\lambda^{(u)}_i$. The gauge-invariant field strengths $K=\frac12K_{\mu\nu}dx_\mu\wedge dx_\nu$ and $\tau_i=\tau_{i\mu}dx_\mu$ read
\begin{equation}
\begin{split}
    K&=*\left(dA^{(\phi)}\right),\\
    \tau_i &= *\left(dA^{(u)}_i-\frac12n_0\epsilon_{ijk}dx_j\wedge dx_k\wedge A^{(\phi)}\right),
\end{split} \label{eq:field_strengths}
\end{equation}
with $*$ denoting the Hodge dual with respect to the Euclidean metric. Physically, $K_{\mu\nu}$ is (the Hodge dual of) the electromagnetic field strength tensor, $\tau_{it}$ is the lattice momentum, and $\tau_{ij}$ is the stress tensor. The currents coupled to the gauge fields are defined as
\begin{equation}
\begin{split}
    J^{\mathrm{vac}}_\mu & = \epsilon_{\mu\nu\gamma\delta}\partial_\nu \partial_\gamma \phi^s_\delta\,,\\
    J^{\mathrm{disl}}_{i\mu\nu} & = \epsilon_{\mu\nu\gamma\delta}\partial_\gamma \partial_\delta u^s_i\,. \label{eq:defects}
\end{split}
\end{equation}
The $\mu=t$ components of the currents in Eq.~(\ref{eq:defects}) are precisely the vacancy density~(\ref{eq:vacancy_density}) and the dislocation density components~(\ref{eq:dislocation_density}). With the above definitions, the full dual action takes the form
\begin{equation}
    S_{\mathrm{dual}}=S_{\mathrm{EM}}+S_{\mathrm{coupling}},
\end{equation}
where
\begin{equation}
\begin{split}
    S_{\mathrm{EM}}=  \frac12 
    \int d^4 x &\left[-c_B^{-1} K_{it}^2 + \frac{c_E^{-1}}{2}K_{ij}^2\right.\\
    &~~\left.-c_{u}^{-1}\tau_{it}^2 +\tilde{C}_{ijkl}\tau_{ij}\tau_{kl}\right]
    \label{eq:SEM}
\end{split}
\end{equation}
with the ``inverse elastic tensor"
\begin{equation}
    \tilde{C}_{ijkl} = \frac{C_b^{-1}}{3} \delta_{ij}\delta_{kl}+\frac{C_s^{-1}}{2}(\delta_{ik}\delta_{jl}+\delta_{il}\delta_{jk}-\frac{2}{3}\delta_{ij}\delta_{kl}),
\end{equation}
and
\begin{equation}
\begin{split}
    S_{\mathrm{coupling}} =   \int d^4 x \left[A^{(\phi)}_\mu J^{\mathrm{vac}}_\mu+\frac12A^{(u)}_{i\mu\nu}J^{\mathrm{disl}}_{i\mu\nu}\right].
\end{split} \label{eq:action_coupling}
\end{equation}
From Eq.~(\ref{eq:SEM}) we can identify 
\begin{equation}
    c_E^{-1}=\varepsilon,\qquad c_B^{-1}=\varepsilon c^2,
\end{equation}
where $\varepsilon$ is the electric permittivity and $c$ is the speed of light.

While the disclination current has not been encountered so far, it does in fact appear in the dual theory as a result of resolving a somewhat subtle issue. The stress tensor as defined in Eq.~(\ref{eq:field_strengths}) contains antisymmetric components, which are unphysical: for this reason, the constraint $\tau_{[ij]}=0$ must be additionally imposed. As shown in App.~\ref{app:duality}, this implies that
\begin{equation}
    \epsilon_{ijk}dx_j\wedge A_k^{(u)}=-\frac{1}{4}d A^{(\Theta)}_i
\label{eq:field_strengths2}
\end{equation}
for some 2-form field $A^{(\Theta)}_i\equiv \frac12 A^{(\Theta)}_{i\mu\nu}dx_\mu \wedge dx_\nu$. While this eliminates the spurious components of $\tau_{ij}$, it also gives rise to one more gauge symmetry, 
\begin{equation}
    A^{(\Theta)}_i\rightarrow A^{(\Theta)}_i+d\lambda^{(\Theta)}_i+4\epsilon_{ijk}dx_j\wedge \lambda^{(u)}_{k}
    \label{eq:disclination_gauge}
\end{equation}
for a 1-form $\lambda^{(\Theta)}_i=\lambda^{(\Theta)}_{i\mu}dx_\mu$. By an explicit evaluation, we find that the action contains a coupling term $\frac{1}{2}A^{(\Theta)}_{i\mu\nu}J^{\mathrm{disc}}_{i\mu\nu}$ with
\begin{equation}
\begin{split}
    J^{\mathrm{disc}}_{i\mu\nu} & = \epsilon_{\mu\nu\gamma\delta}\partial_\gamma \partial_\delta \Theta^s_i \\
    & = \frac12 \epsilon_{ijk}\partial_jJ^{\mathrm{disl}}_{k\mu\nu}.
    \label{eq:disclinations}
\end{split}
\end{equation}
The final equality in Eq.~(\ref{eq:disclinations}) proves that the disclination current can be obtained as a curl of the dislocation current, and is therefore not an independent quantity. This equality can also be used to show that a single disclination is equivalent to a finite-density texture of dislocations, as suggested earlier in our discussion of disclination defects.

Important insights can be gained by imposing the gauge invariance~(\ref{eq:gauge}),~(\ref{eq:disclination_gauge}) on the action~(\ref{eq:action_coupling}). This leads to a set of continuity equations
\begin{subequations}
\begin{align}
        \begin{split}
        &\partial_t\left(\rho^{\mathrm{vac}}-n_0\epsilon_{ijk}x_i\rho^{\mathrm{disl}}_{jk}\right)\qquad\qquad\qquad \\&\qquad\qquad+\partial_i\left(J^{\mathrm{vac}}_i-n_0\epsilon_{jkl}x_jJ^{\mathrm{disl}}_{kli}\right)=0,\end{split}\label{eq:cont3}\\
        &\partial_\nu J^{\mathrm{disc}}_{i\mu\nu}=0,\label{eq:cont1}\\
        &\partial_\nu J^{\mathrm{disl}}_{i\mu\nu}=0.\label{eq:cont2}
\end{align} \label{eq:cont}
\end{subequations}
The following facts can be deduced from Eqs.~(\ref{eq:cont}).
\begin{itemize}
    \item Integrating the $\mu=t$ component of Eq.~(\ref{eq:cont1}) ((\ref{eq:cont2})) over a finite volume, we find that the total Frank vector (Burgers vector) of defects crossing any closed compact surface is zero. Thus, disclinations (dislocations) must form closed loops or extend to the boundaries of the system.
    \item Calculating the surface integral of Eq.~(\ref{eq:cont1}) ((\ref{eq:cont2})) over a closed spatial surface, we find that the total Frank vector (Burgers vector) of defects crossing any closed surface is conserved. For example, assuming that the auxiliary surface in Fig.~\ref{fig:dislocation} extends to infinity, its total Frank vector and Burgers vector remain constant in time.
    \item Eq.~(\ref{eq:cont3}) expresses the conservation of the total electric charge. It implies that a dislocation line is free to move in the plane spanned by the Burgers vector and the direction of the dislocation line $\boldsymbol\rho^{\mathrm{disl}}_{i}$, motion known in the theory of elasticity as a glide. On the other hand, motion perpendicular to that plane, known as a climb, has to be accompanied by absorbing or emitting a vacancy.
\end{itemize}
All of the above-described phenomenology agrees very well with the behavior of defects observed in uncharged crystals. The above conservation laws can be more succinctly expressed as conservation of the charges:
\begin{subequations}
\begin{align}
    Q^{\mathrm{disc}}_i[\Sigma] & = \int_\Sigma *J^{\mathrm{disc}}_i,\\
    Q^{\mathrm{disl}}_i[\Sigma] & = \int_\Sigma *J^{\mathrm{disl}}_i,\\
    Q^{\mathrm{vac}}&=\int d^3 x \left(\rho^{\mathrm{vac}}+n_0\sum_{i=1}^3\left(\vct x \times\boldsymbol\rho^{\mathrm{disl}}_{i}\right)\cdot \frac{b_i}{|b|}\right),
\end{align}
\end{subequations}
where $\Sigma$ is an arbitrary closed two-dimensional hypersurface, $b_i$ is the components of the Burgers vector associated with the given dislocation line and $|b|$ is the length of the Burgers vector. 

\section{Interactions of vacancies and interstitials}
\label{sec:screening}

As the first application of the gauge theory developed in the previous sections, we find the electric field and the stress fields produced by vacancies and interstitials, and calculate their interaction energy. We will be looking for a static configuration of the fields $K$ and $\tau_i$, meaning we set all the time derivatives and defect currents to zero: $\partial_t(\ldots)=0$, $J^{\mathrm{vac}}_i=J^{\mathrm{disl}}_{ijk}=0$. Defining $E_i = \frac12\epsilon_{ijk}K_{jk}$, we find the following set of equations of motion and Bianchi identities:
\begin{subequations}
\begin{align}
    \varepsilon \partial_iE_i+n_0C_b^{-1}\tau_{ii}&=\rho^{\mathrm{vac}}, \label{eq:field_vac}\\
    \epsilon_{ijk}\partial_j E_k &=0,\\
    \epsilon_{jkl}\partial_k u_{il} &=\rho^{\mathrm{disl}}_{ij},\\
    \partial_j\tau_{ij}&=-n_0E_i,
\end{align} \label{eq:field_eqs}
\end{subequations}
where the relation between the stress and the strain is
\begin{equation}
    \tau_{ij} = C_{ijkl}u_{kl}.
\end{equation}
Here, in order to find the fields produced by a point charge, we set
\begin{equation}
    \rho^{\mathrm{vac}}=Q\delta^3(\vct x),\qquad \rho^{\mathrm{disl}}_{ij}=0.
\end{equation}
In the static case, the electric field is given by the electrostatic potential through $E_i = -\partial_i V$. Using Eqs.~(\ref{eq:field_eqs}) and the spherical symmetry, we find 
\begin{equation}
    V(r) = \frac{Q}{4 \pi\varepsilon}\frac{e^{-r/l}}{r},
\end{equation}
where the screening length is
\begin{equation}
    l = \sqrt{\frac{\varepsilon}{n_0^2}\frac{2C_s+C_b}{3}}.
\end{equation}
Using the values for the elastic constants given in~\cite{Tozzini_1996}, in which the authors computed the elastic constants in body-centered-cubic Wigner crystals near melting, we find
\begin{equation}
    l\approx0.25n_0^{-1/3}.
\end{equation}
In fact, while~\cite{Tozzini_1996} only studies crystals near melting, the observation that on the microscopic level there is no length scale other than $n_0^{-1/3}$ implies that we can generically expect $l\sim n_0^{-1/3}$; in other words, the screening length is of the order of the average electron-electron distance. On such a distance, however, the formalism developed here to study the long-wavelength behavior of the crystal lattice loses its validity. 
Thus, the displacement of lattice sites around a vacancy or an interstitial must be calculated from a microscopic model, like in \cite{Fisher1979,Cockayne1990,Candido2001}.

On the other hand, while there is no long-range electrostatic potential, the displacement vector in the radial direction is
\begin{equation}
    u_r(r) = n_0^{-1}\frac{Q}{4 \pi r^2}+\mathcal{O}(e^{-r/l}).
\end{equation}
With such a displacement field, the total charge enclosed by any sphere centered on the defect is zero. While far from the defect the lattice compression $\partial_iu_i$ is zero, the shear strain $\partial_{\langle i}u_{j\rangle}$ falls off as $r^{-3}$. In the presence of two point defects with charges $Q_1$ and $Q_2$ separated by a distance $R$, the interaction energy calculated as the difference between the elastic energy of the composite system minus the elastic energy of both defects in isolation is
\begin{equation}
    E_{\mathrm{int}}=\frac{C_s}{2 \pi n_0^2}\frac{Q_1Q_2}{R^3}.
\end{equation}
This result implies that long-range intraspecies interaction is repulsive and interspecies interaction is attractive. This stands in contrast to the result of \cite{Candido2001}, which found that vacancy-vacancy interaction is strongly attractive at least up to 8 lattice sites apart, and to \cite{Fisher1979}, which found that a divacancy has a much smaller energy than two isolated vacancies. On the other hand, the above-mentioned papers only studied short-range physics, which is the opposite limit to the one considered here.

\section{Condensation of vacancies or interstitials}
\label{sec:higgs}

The realization that elastic defects can be treated as matter currents coupled to dynamical gauge fields allows for an analytical approach to the collective physics of defects. In this section, inspired by suggestions that in a certain parameter regime interstitials might proliferate \cite{Barraza1999,Pankov2008,Kivelson2024}, we study the implications of a formation of a condensate of vacancies or interstitials. Since in an electron crystal the point defects in question are fermions, this would in principle require a pairing mechanism between them; however, we stress that we do not speculate on physical mechanisms than can lead to the charge condensation, only focusing on its consequences. Furthermore, some of the phenomena that we find can be generically expected whenever the Wigner crystal has a finite ground-state density of vacancies and interstitials, as shall become clearer when we discuss the physical intuition behind our results.

In what follows, we assume the absence of dislocations. The starting point of the analysis is the Ginzburg-Landau-type functional
\begin{equation}
    S_{\mathrm{GL}} = S_{\mathrm{EM}}+S_{\mathrm{vac}},
\end{equation}
where $S_{\mathrm{EM}}$ is given in Eq.~(\ref{eq:SEM}) and $S_{\mathrm{vac}}$ describes the dynamics of the order parameter $\psi$ representing vacancies:
\begin{multline}
    S_{\mathrm{vac}} = -\int d^4 x \left[\psi^{\dagger}\left(i\partial_t+A^{(\phi)}_t\right)\psi\right.\\~\left.+\frac{1}{2m^*}\left|\left(i\partial_i+A^{(\phi)}_i\right)\psi\right|^2
    +\alpha |\psi|^2+\frac{\beta}{2}|\psi|^4\right].
    \label{eq:action_vacancy}
\end{multline}
In Eq.~(\ref{eq:action_vacancy}), $\psi$ is a scalar Higgs field transforming linearly under $U(1)$, $m^*$ is the effective mass, and as usual the parameter $\beta$ is positive while $\alpha$ changes sign at the critical point. In the symmetric phase, where $\alpha>0$, the minimum of the potential is at $\psi=0$ and the Higgs field is gapped. In the Higgs phase, where $\alpha<0$, the minimum of the potential is at $|\psi|_0^2 =\frac{\beta}{-\alpha}$ and the phase degree of freedom of $\psi$ becomes gapless, while the amplitude mode remains gapped. In other words, in the decomposition $\psi=\left(|\psi|_0+\delta|\psi|\right)e^{i\theta}$, $\delta|\psi|$ is a massive mode that can be integrated out at low energies, leaving the phase $\theta$ as the only relevant degree of freedom. After integrating out $\delta|\psi|$, the leading-order terms are
\begin{multline}
    S_{\mathrm{vac}} \rightarrow S_{\theta} = \rho_s\int d^4 x \left[m^*\left(\partial_t\theta-A^{(\phi)}_t\right)\right.\\
    \left.-\frac{1}{2}\left(\partial_i\theta-A^{(\phi)}_i\right)^2  +\frac{1}{2c^{*2}}\left(\partial_t\theta-A^{(\phi)}_t\right)^2\right],
\end{multline}
where $\rho_s=\frac{|\psi|_0^2}{m^*}$ is the superfluid stiffness and we have defined the quantity $c^*=\sqrt{\beta\rho_s}$ with dimensions of velocity.
Since we are only interested in studying the low-wavelength behavior, we can write in general
\begin{equation}
    S = S_{\mathrm{EM}}+S_{\theta},
    \label{eq:action_Higgs}
\end{equation}
with the understanding that $\rho_s=0$ in the symmetric phase.

Mode dispersions can be found after deriving the equations of motion from the action~(\ref{eq:action_Higgs}). In App.~\ref{app:higgs}, we pursue another approach, in which the remaining gapped modes are first integrated out, revealing more directly the low-energy structure of the effective action. Either way, the massless spectrum is found to be as follows.
\begin{itemize}
    \item When $\rho_s=0$, longitudinal oscillations are gapped and transverse oscillations exhibit a quadratic dispersion. For a wave traveling in the $x$ direction, we obtain
    \begin{equation}
        \tau_{\langle xy\rangle},~\tau_{\langle xz\rangle}: \qquad\omega = \pm \sqrt{\frac{C_s}{2n_0^2 c_B}}k^2.
    \end{equation}
    This is consistent with Eq.~(\ref{eq:dispersion_unbroken}).
    \item When $\rho_s\neq0$, longitudinal oscillations are permitted and all the modes are linear. For a wave traveling in the $x$ direction, we obtain
    \begin{equation}
    \begin{split}
     \tau_{\langle xy\rangle},~\tau_{\langle xz\rangle}:\qquad& \omega = \pm \sqrt{\frac{C_s\rho_s}{2\left(n_0^2+c_u\rho_s\right)}}k.\\
    \frac13\tau_{ii}-\left(\partial_t\theta-A^{(\phi)}_t\right):\qquad& \omega =\pm \sqrt{\frac{3 n_0^2 c^{*2}+C_b \rho_s}{3 n_0^2 +c_u \rho_s}}k.
    \end{split}
    \end{equation}
\end{itemize}
These results suggest the following behavior of the longitudinal mode. In the Coulomb phase, longitudinal oscillations have a gap equal to the plasma frequency, as displacing the lattice electrons (as in Fig.~\ref{fig:displacement}) results in charge separation, leading to finite-frequency oscillations. In this phase, lattice compressions are not allowed at low frequencies. On the other hand, in the presence of a finite density of point defects, one can imagine periodic lattice compressions accompanied by a redistribution of the vacancies/interstitials in such a way that the coarse-grained charged density remains zero everywhere. This way, plasma oscillations are avoided and long-wavelength longitudinal waves can be present. Based on this intuition, we speculate that the appearance of a longitudinal mode is a generic scenario for charged crystals in the presence of a finite density of mobile charged defects. At the same time, in the extreme case of a fully melted crystal, all the modes are gapped by plasma oscillations \cite{Vlasov_1968}, meaning that a coexistence of the elastic lattice and the charged defects is crucial for the presence of gapless sound waves.

Note that in the scenario discussed in the section, spontaneous breaking of a gauge symmetry increases the number of gapless modes in the system instead of decreasing it, at odds with the usual Higgs mechanism. Such a phenomenon was first discussed in Ref. \cite{glodkowski2025}. From the formal point of view it can be attributed to the fact that the gauge transformations do not commute with translations, which leads to the situation in which some of the invariant field strengths contain gauge fields without derivatives. This in turn makes these field strengths massive even in the unbroken-symmetry case \cite{Hirono_2024}. In our case these massive field strengths contain $A^{(\phi)}_\mu$, see Eq.~(\ref{eq:field_strengths}). As a result, when matter coupled to $A^{(\phi)}_\mu$ undergoes condensation, the usual Higgs mechanism does not take place.

The scenario discussed here is qualitatively different from the vacancy condensation in superfluid vortex crystals analyzed in \cite{Nguyen_2020}. In the latter case, the breaking of time-reversal symmetry by the background magnetic field introduces a gyroscopic term proportional to $\epsilon_{ij}u_i\partial_tu_j$ in the (2+1)-dimensional Lagrangian. This term, which has no counterpart in our action~(\ref{eq:action}), renders $u_x$ and $u_y$ as a pair of canonically conjugate variables. Consequently, the vortex crystal supports only a single gapless mode with a quadratic dispersion, preventing the anti-Higgs mechanism from occurring.

\section{Discussion}

In this paper, we have developed an effective field theory of a charged crystal with defects. The different species of defects -- namely, vacancies/interstitials, dislocations, and disclinations -- are represented as particles interacting with a set of gauge fields corresponding to the elastic and electromagnetic degrees of freedom. The unusual form of gauge transformations is shown to determine both the interplay between the elasticity and electromagnetism, and the fractonic behavior of dislocations and disclinations. Then, the effective field theory is used to determine the vacancy-vacancy interaction energy, and to find that condensation of charged defects leads to the appearance of an extra longitudinal gapless mode via the anti-Higgs mechanism.

While we have focused on the applications of our theory to describe the physics of vacancies/interstitials, it can equally well be applied to systems with dislocations or disclinations. It can also be used as a starting point in the calculation of finite-temperature transport properties of charged crystals.

There are various ways in which the present theory can be adapted to better describe certain realistic systems. Firstly, we have neglected the fermionic statistics of the electron, as well as the spin degree of freedom, which endows real Wigner crystals with magnetic properties. Secondly, in real systems the background explicitly breaks translational symmetry, pinning the crystals and gapping out some modes \cite{Chitra_2005, Delacretaz2017,Lier2023}. The theory of charged crystal elasticity can also potentially be adapted to systems other than three-dimensional Wigner crystals, such as two-dimensional Wigner crystals, bilayer Wigner crystals \cite{Zhou_2021,Zhuang2025}, ion Coulomb crystals \cite{Schneider_2012,Thompson_2015}, or generalized Wigner crystals \cite{Regan2020,Reichhardt2025}.

\section{Acknowledgements}

I thank Tobias Holder, Francisco Pe\~na-Ben\'\i{}tez, Aleksander Głódkowski and Lazaros Tsaloukidis for comments on the manuscript. I acknowledge financial support by the European Research Council (ERC) under grant QuantumCUSP (Grant Agreement No. 101077020) and by the Deutsche Forschungsgemein-
schaft (DFG) through the cluster of excellence ct.qmat (EXC
2147, project-id 390858490).

\bibliography{quadrupole}

@Article{Wigner1938,
author ="Wigner, E.",
title  ="Effects of the electron interaction on the energy levels of electrons in metals",
journal  ="Trans. Faraday Soc. ",
year  ="1938",
volume  ="34",
issue  ="0",
pages  ="678-685",
publisher  ="The Royal Society of Chemistry",
doi  ="10.1039/TF9383400678",
url  ="http://dx.doi.org/10.1039/TF9383400678"}

@article{Fuchs1935,
author = {Fuchs, K. },
title = {A quantum mechanical investigation of the cohesive forces of metallic copper},
journal = {Proceedings of the Royal Society of London. Series A - Mathematical and Physical Sciences},
volume = {151},
number = {874},
pages = {585-602},
year = {1935},
doi = {10.1098/rspa.1935.0167},
URL = {https://royalsocietypublishing.org/doi/abs/10.1098/rspa.1935.0167}
}

@article{Bonsall1977,
  title = {Some static and dynamical properties of a two-dimensional Wigner crystal},
  author = {Bonsall, Lynn and Maradudin, A. A.},
  journal = {Phys. Rev. B},
  volume = {15},
  issue = {4},
  pages = {1959--1973},
  numpages = {0},
  year = {1977},
  month = {Feb},
  publisher = {American Physical Society},
  doi = {10.1103/PhysRevB.15.1959},
  url = {https://link.aps.org/doi/10.1103/PhysRevB.15.1959}
}

@article{Ceperley1980,
  title = {Ground State of the Electron Gas by a Stochastic Method},
  author = {Ceperley, D. M. and Alder, B. J.},
  journal = {Phys. Rev. Lett.},
  volume = {45},
  issue = {7},
  pages = {566--569},
  numpages = {0},
  year = {1980},
  month = {Aug},
  publisher = {American Physical Society},
  doi = {10.1103/PhysRevLett.45.566},
  url = {https://link.aps.org/doi/10.1103/PhysRevLett.45.566}
}

@article{Ceperley1989,
  title = {Ground state of the two-dimensional electron gas},
  author = {Tanatar, B. and Ceperley, D. M.},
  journal = {Phys. Rev. B},
  volume = {39},
  issue = {8},
  pages = {5005--5016},
  numpages = {0},
  year = {1989},
  month = {Mar},
  publisher = {American Physical Society},
  doi = {10.1103/PhysRevB.39.5005},
  url = {https://link.aps.org/doi/10.1103/PhysRevB.39.5005}
}

@article{Huang_2021,
   title={Correlated insulating states at fractional fillings of the WS2/WSe2 moiré lattice},
   volume={17},
   ISSN={1745-2481},
   url={http://dx.doi.org/10.1038/s41567-021-01171-w},
   DOI={10.1038/s41567-021-01171-w},
   number={6},
   journal={Nature Physics},
   publisher={Springer Science and Business Media LLC},
   author={Huang, Xiong and Wang, Tianmeng and Miao, Shengnan and Wang, Chong and Li, Zhipeng and Lian, Zhen and Taniguchi, Takashi and Watanabe, Kenji and Okamoto, Satoshi and Xiao, Di and Shi, Su-Fei and Cui, Yong-Tao},
   year={2021},
   month=feb, pages={715–719} }

@article{Zhou_2021,
   title={Bilayer Wigner crystals in a transition metal dichalcogenide heterostructure},
   volume={595},
   ISSN={1476-4687},
   url={http://dx.doi.org/10.1038/s41586-021-03560-w},
   DOI={10.1038/s41586-021-03560-w},
   number={7865},
   journal={Nature},
   publisher={Springer Science and Business Media LLC},
   author={Zhou, You and Sung, Jiho and Brutschea, Elise and Esterlis, Ilya and Wang, Yao and Scuri, Giovanni and Gelly, Ryan J. and Heo, Hoseok and Taniguchi, Takashi and Watanabe, Kenji and Zaránd, Gergely and Lukin, Mikhail D. and Kim, Philip and Demler, Eugene and Park, Hongkun},
   year={2021},
   month=jun, pages={48–52} }

@article{Smolenski_2021,
   title={Signatures of Wigner crystal of electrons in a monolayer semiconductor},
   volume={595},
   ISSN={1476-4687},
   url={http://dx.doi.org/10.1038/s41586-021-03590-4},
   DOI={10.1038/s41586-021-03590-4},
   number={7865},
   journal={Nature},
   publisher={Springer Science and Business Media LLC},
   author={Smoleński, Tomasz and Dolgirev, Pavel E. and Kuhlenkamp, Clemens and Popert, Alexander and Shimazaki, Yuya and Back, Patrick and Lu, Xiaobo and Kroner, Martin and Watanabe, Kenji and Taniguchi, Takashi and Esterlis, Ilya and Demler, Eugene and Imamoğlu, Ataç},
   year={2021},
   month=jun, pages={53–57} }

@article{Thomas1966,
author = {T. Y. Thomas },
title = {ON THE STRESS-STRAIN RELATIONS FOR CUBIC CRYSTALS},
journal = {Proceedings of the National Academy of Sciences},
volume = {55},
number = {2},
pages = {235-239},
year = {1966},
doi = {10.1073/pnas.55.2.235},
URL = {https://www.pnas.org/doi/abs/10.1073/pnas.55.2.235},
eprint = {https://www.pnas.org/doi/pdf/10.1073/pnas.55.2.235}}

@article{Tozzini_1996,
doi = {10.1088/0953-8984/8/43/009},
url = {https://dx.doi.org/10.1088/0953-8984/8/43/009},
year = {1996},
month = {oct},
publisher = {},
volume = {8},
number = {43},
pages = {8121},
author = {V Tozzini and M P Tosi},
title = {Lattice vibrations and elastic constants of three- and two-dimensional quantal Wigner crystals near melting},
journal = {Journal of Physics: Condensed Matter}
}

@article{Fisher1979,
  title = {Defects in the two-dimensional electron solid and implications for melting},
  author = {Fisher, Daniel S. and Halperin, B. I. and Morf, R.},
  journal = {Phys. Rev. B},
  volume = {20},
  issue = {11},
  pages = {4692--4712},
  numpages = {0},
  year = {1979},
  month = {Dec},
  publisher = {American Physical Society},
  doi = {10.1103/PhysRevB.20.4692},
  url = {https://link.aps.org/doi/10.1103/PhysRevB.20.4692}
}

@article{Cockayne1990,
  title = {Energetics of point defects in the two-dimensional Wigner crystal},
  author = {Cockayne, Eric and Elser, Veit},
  journal = {Phys. Rev. B},
  volume = {43},
  issue = {1},
  pages = {623--629},
  numpages = {0},
  year = {1991},
  month = {Jan},
  publisher = {American Physical Society},
  doi = {10.1103/PhysRevB.43.623},
  url = {https://link.aps.org/doi/10.1103/PhysRevB.43.623}
}

@article{Barraza1999,
title = {Vacancies in quantal Wigner crystals near melting},
journal = {Solid State Communications},
volume = {112},
number = {5},
pages = {261-264},
year = {1999},
issn = {0038-1098},
doi = {https://doi.org/10.1016/S0038-1098(99)00341-5},
url = {https://www.sciencedirect.com/science/article/pii/S0038109899003415},
author = {N. Barraza and L. Colletti and M.P. Tosi}
}

@article{Candido2001,
  title = {Single and Paired Point Defects in a 2D Wigner Crystal},
  author = {C\^andido, Ladir and Phillips, Philip and Ceperley, D. M.},
  journal = {Phys. Rev. Lett.},
  volume = {86},
  issue = {3},
  pages = {492--495},
  numpages = {0},
  year = {2001},
  month = {Jan},
  publisher = {American Physical Society},
  doi = {10.1103/PhysRevLett.86.492},
  url = {https://link.aps.org/doi/10.1103/PhysRevLett.86.492}
}

@article{Pankov2008,
  title = {Self-doping instability of the Wigner-Mott insulator},
  author = {Pankov, S. and Dobrosavljevi\ifmmode \acute{c}\else \'{c}\fi{}, V.},
  journal = {Phys. Rev. B},
  volume = {77},
  issue = {8},
  pages = {085104},
  numpages = {5},
  year = {2008},
  month = {Feb},
  publisher = {American Physical Society},
  doi = {10.1103/PhysRevB.77.085104},
  url = {https://link.aps.org/doi/10.1103/PhysRevB.77.085104}
}

@article{Grimes_1980,
title = {Crystallization of electrons on the surface of liquid helium},
journal = {Surface Science},
volume = {98},
number = {1},
pages = {1-7},
year = {1980},
issn = {0039-6028},
doi = {https://doi.org/10.1016/0039-6028(80)90465-3},
url = {https://www.sciencedirect.com/science/article/pii/0039602880904653},
author = {C.C. Grimes and G. Adams}
}

@article{Yoon1999,
  title = {Wigner Crystallization and Metal-Insulator Transition of Two-Dimensional Holes in GaAs at $\mathit{B}\phantom{\rule{0ex}{0ex}}=\phantom{\rule{0ex}{0ex}}0$},
  author = {Yoon, Jongsoo and Li, C. C. and Shahar, D. and Tsui, D. C. and Shayegan, M.},
  journal = {Phys. Rev. Lett.},
  volume = {82},
  issue = {8},
  pages = {1744--1747},
  numpages = {0},
  year = {1999},
  month = {Feb},
  publisher = {American Physical Society},
  doi = {10.1103/PhysRevLett.82.1744},
  url = {https://link.aps.org/doi/10.1103/PhysRevLett.82.1744}
}

@inproceedings{Huang_2014,
   title={Experimental study of two-dimensional quantum Wigner solid in zero magnetic field},
   ISSN={0094-243X},
   url={http://dx.doi.org/10.1063/1.4870193},
   DOI={10.1063/1.4870193},
   booktitle={AIP Conference Proceedings},
   publisher={AIP Publishing LLC},
   author={Huang, Jian and Pfeiffer, L. N. and West, K. W.},
   year={2014},
   pages={42–50} }

@Article{Sung2025,
author={Sung, Jiho
and Wang, Jue
and Esterlis, Ilya
and Volkov, Pavel A.
and Scuri, Giovanni
and Zhou, You
and Brutschea, Elise
and Taniguchi, Takashi
and Watanabe, Kenji
and Yang, Yubo
and Morales, Miguel A.
and Zhang, Shiwei
and Millis, Andrew J.
and Lukin, Mikhail D.
and Kim, Philip
and Demler, Eugene
and Park, Hongkun},
title={An electronic microemulsion phase emerging from a quantum crystal-to-liquid transition},
journal={Nature Physics},
year={2025},
month={Mar},
day={01},
volume={21},
number={3},
pages={437-443},
issn={1745-2481},
doi={10.1038/s41567-024-02759-8},
url={https://doi.org/10.1038/s41567-024-02759-8}
}

@article{Chitra_2005,
   title={Zero field Wigner crystal},
   volume={44},
   ISSN={1434-6036},
   url={http://dx.doi.org/10.1140/epjb/e2005-00145-0},
   DOI={10.1140/epjb/e2005-00145-0},
   number={4},
   journal={The European Physical Journal B},
   publisher={Springer Science and Business Media LLC},
   author={Chitra, R. and Giamarchi, T.},
   year={2005},
   month=apr, pages={455–467} }

@article{Delacretaz2017,
  title = {Theory of hydrodynamic transport in fluctuating electronic charge density wave states},
  author = {Delacr\'etaz, Luca V. and Gout\'eraux, Blaise and Hartnoll, Sean A. and Karlsson, Anna},
  journal = {Phys. Rev. B},
  volume = {96},
  issue = {19},
  pages = {195128},
  numpages = {17},
  year = {2017},
  month = {Nov},
  publisher = {American Physical Society},
  doi = {10.1103/PhysRevB.96.195128},
  url = {https://link.aps.org/doi/10.1103/PhysRevB.96.195128}
}

@article{Lier2023,
  title = {Approximate symmetries, pseudo-Goldstones, and the second law of thermodynamics},
  author = {Armas, Jay and Jain, Akash and Lier, Ruben},
  journal = {Phys. Rev. D},
  volume = {108},
  issue = {8},
  pages = {086011},
  numpages = {13},
  year = {2023},
  month = {Oct},
  publisher = {American Physical Society},
  doi = {10.1103/PhysRevD.108.086011},
  url = {https://link.aps.org/doi/10.1103/PhysRevD.108.086011}
}

@article{Vlasov_1968,
doi = {10.1070/PU1968v010n06ABEH003709},
url = {https://dx.doi.org/10.1070/PU1968v010n06ABEH003709},
year = {1968},
month = {jun},
publisher = {},
volume = {10},
number = {6},
pages = {721},
author = {A A Vlasov},
title = {THE VIBRATIONAL PROPERTIES OF AN ELECTRON GAS},
journal = {Soviet Physics Uspekhi}
}

@article{Zhuang2025,
  title = {Defect liquids in a weakly imbalanced bilayer Wigner crystal},
  author = {Zhuang, Zekun and Esterlis, Ilya},
  journal = {Phys. Rev. B},
  volume = {111},
  issue = {20},
  pages = {205122},
  numpages = {8},
  year = {2025},
  month = {May},
  publisher = {American Physical Society},
  doi = {10.1103/PhysRevB.111.205122},
  url = {https://link.aps.org/doi/10.1103/PhysRevB.111.205122}
}

@article{Schneider_2012,
doi = {10.1088/0034-4885/75/2/024401},
url = {https://dx.doi.org/10.1088/0034-4885/75/2/024401},
year = {2012},
month = {jan},
publisher = {},
volume = {75},
number = {2},
pages = {024401},
author = {Schneider, Ch and Porras, Diego and Schaetz, Tobias},
title = {Experimental quantum simulations of many-body physics with trapped ions},
journal = {Reports on Progress in Physics}
}

@article{Thompson_2015,
   title={Ion Coulomb crystals},
   volume={56},
   ISSN={1366-5812},
   url={http://dx.doi.org/10.1080/00107514.2014.989715},
   DOI={10.1080/00107514.2014.989715},
   number={1},
   journal={Contemporary Physics},
   publisher={Informa UK Limited},
   author={Thompson, Richard C.},
   year={2015},
   month=jan, pages={63–79} }

@Article{Regan2020,
author={Regan, Emma C.
and Wang, Danqing
and Jin, Chenhao
and Bakti Utama, M. Iqbal
and Gao, Beini
and Wei, Xin
and Zhao, Sihan
and Zhao, Wenyu
and Zhang, Zuocheng
and Yumigeta, Kentaro
and Blei, Mark
and Carlstr{\"o}m, Johan D.
and Watanabe, Kenji
and Taniguchi, Takashi
and Tongay, Sefaattin
and Crommie, Michael
and Zettl, Alex
and Wang, Feng},
title={Mott and generalized Wigner crystal states in WSe2/WS2 moir{\'e} superlattices},
journal={Nature},
year={2020},
month={Mar},
day={01},
volume={579},
number={7799},
pages={359-363},
issn={1476-4687},
doi={10.1038/s41586-020-2092-4},
url={https://doi.org/10.1038/s41586-020-2092-4}
}

@article{Reichhardt2025,
  title = {Depinning, melting, and sliding of generalized Wigner crystals in Moir\'e systems},
  author = {Reichhardt, C. and Reichhardt, C. J. O.},
  journal = {Phys. Rev. Res.},
  volume = {7},
  issue = {1},
  pages = {013155},
  numpages = {14},
  year = {2025},
  month = {Feb},
  publisher = {American Physical Society},
  doi = {10.1103/PhysRevResearch.7.013155},
  url = {https://link.aps.org/doi/10.1103/PhysRevResearch.7.013155}
}

@article{Gaiotto_2015,
   title={Generalized global symmetries},
   volume={2015},
   ISSN={1029-8479},
   url={http://dx.doi.org/10.1007/JHEP02(2015)172},
   DOI={10.1007/jhep02(2015)172},
   number={2},
    pages={172},
   journal={Journal of High Energy Physics},
   publisher={Springer Science and Business Media LLC},
   author={Gaiotto, Davide and Kapustin, Anton and Seiberg, Nathan and Willett, Brian},
   year={2015},
   month=feb }

@article{Gomes_2023,
   title={An introduction to higher-form symmetries},
   ISSN={2590-1990},
   url={http://dx.doi.org/10.21468/SciPostPhysLectNotes.74},
   DOI={10.21468/scipostphyslectnotes.74},
   journal={SciPost Physics Lecture Notes},
    pages={74},
   publisher={Stichting SciPost},
   author={Gomes, Pedro R. S.},
   year={2023},
   month=sep }

@article{Kleinert1982,
	author = {H. Kleinert},
	doi = {10.1016/0375-9601(82)90578-3},
	issn = {0375-9601},
	journal = {Physics Letters A},
	number = {6},
	pages = {295–298},
	title = {Duality transformation for defect melting},
	url = {https://www.sciencedirect.com/science/article/pii/0375960182905783},
	volume = {91},
	year = {1982}
}

@article{Kleinert1983,
	author = {H. Kleinert},
	doi = {10.1016/0375-9601(83)90185-8},
	issn = {0375-9601},
	journal = {Physics Letters A},
	number = {6},
	pages = {302–306},
	title = {Dual model for dislocation and disclination melting},
	url = {https://www.sciencedirect.com/science/article/pii/0375960183901858},
	volume = {96},
	year = {1983}
}

@article{Zaanen_2004,
   title={Duality in 2+1D quantum elasticity: superconductivity and quantum nematic order},
   volume={310},
   ISSN={0003-4916},
   url={http://dx.doi.org/10.1016/j.aop.2003.10.003},
   DOI={10.1016/j.aop.2003.10.003},
   number={1},
   journal={Annals of Physics},
   publisher={Elsevier BV},
   author={Zaanen, J. and Nussinov, Z. and Mukhin, S.I.},
   year={2004},
   month=mar, pages={181–260} }

@article{Cvetkovic_2006,
   title={Topological kinematic constraints: dislocations and the glide principle},
   volume={86},
   ISSN={1478-6443},
   url={http://dx.doi.org/10.1080/14786430600636328},
   DOI={10.1080/14786430600636328},
   number={20},
   journal={Philosophical Magazine},
   publisher={Informa UK Limited},
   author={Cvetkovic, V. and Nussinov, Z. and Zaanen, J.},
   year={2006},
   month=jul, pages={2995–3020} }

@article{Beekman_2017a,
   title={Dual gauge field theory of quantum liquid crystals in two dimensions},
   volume={683},
   ISSN={0370-1573},
   url={http://dx.doi.org/10.1016/j.physrep.2017.03.004},
   DOI={10.1016/j.physrep.2017.03.004},
   journal={Physics Reports},
   publisher={Elsevier BV},
   author={Beekman, Aron J. and Nissinen, Jaakko and Wu, Kai and Liu, Ke and Slager, Robert-Jan and Nussinov, Zohar and Cvetkovic, Vladimir and Zaanen, Jan},
   year={2017},
   month=apr, pages={1–110} }

@article{Beekman_2017b,
  title = {Dual gauge field theory of quantum liquid crystals in three dimensions},
  author = {Beekman, Aron J. and Nissinen, Jaakko and Wu, Kai and Zaanen, Jan},
  journal = {Phys. Rev. B},
  volume = {96},
  issue = {16},
  pages = {165115},
  numpages = {50},
  year = {2017},
  month = {Oct},
  publisher = {American Physical Society},
  doi = {10.1103/PhysRevB.96.165115},
  url = {https://link.aps.org/doi/10.1103/PhysRevB.96.165115}
}

@article{Mermin1979,
  title = {The topological theory of defects in ordered media},
  author = {Mermin, N. D.},
  journal = {Rev. Mod. Phys.},
  volume = {51},
  issue = {3},
  pages = {591--648},
  numpages = {0},
  year = {1979},
  month = {Jul},
  publisher = {American Physical Society},
  doi = {10.1103/RevModPhys.51.591},
  url = {https://link.aps.org/doi/10.1103/RevModPhys.51.591}
}

@book{Kleinert1989,
    author = "Kleinert, Hagen",
    title = "{Gauge Fields in Condensed Matter}",
    volume = "II",
    doi = "10.1142/0356",
    isbn = "978-9971-5-0210-2",
    publisher = "World Scientific",
    address = "Singapore",
    year = "1989"
}

@article{Pretko2018a,
  title = {Fracton-Elasticity Duality},
  author = {Pretko, Michael and Radzihovsky, Leo},
  journal = {Phys. Rev. Lett.},
  volume = {120},
  issue = {19},
  pages = {195301},
  numpages = {7},
  year = {2018},
  month = {May},
  publisher = {American Physical Society},
  doi = {10.1103/PhysRevLett.120.195301},
  url = {https://link.aps.org/doi/10.1103/PhysRevLett.120.195301}
}

@article{Pretko:2018jbi,
    author = "Pretko, Michael",
    title = "{The Fracton Gauge Principle}",
    archivePrefix = "arXiv",
    primaryClass = "cond-mat.str-el",
    doi = "10.1103/PhysRevB.98.115134",
    journal = "Phys. Rev. B",
    volume = "98",
    number = "11",
    pages = "115134",
    year = "2018"
}

@article{Pretko2018b,
  title = {Fractonic line excitations: An inroad from three-dimensional elasticity theory},
  author = {Pai, Shriya and Pretko, Michael},
  journal = {Phys. Rev. B},
  volume = {97},
  issue = {23},
  pages = {235102},
  numpages = {8},
  year = {2018},
  month = {Jun},
  publisher = {American Physical Society},
  doi = {10.1103/PhysRevB.97.235102},
  url = {https://link.aps.org/doi/10.1103/PhysRevB.97.235102}
}

@article{Pretko2019,
  title = {Crystal-to-fracton tensor gauge theory dualities},
  author = {Pretko, Michael and Zhai, Zhengzheng and Radzihovsky, Leo},
  journal = {Phys. Rev. B},
  volume = {100},
  issue = {13},
  pages = {134113},
  numpages = {33},
  year = {2019},
  month = {Oct},
  publisher = {American Physical Society},
  doi = {10.1103/PhysRevB.100.134113},
  url = {https://link.aps.org/doi/10.1103/PhysRevB.100.134113}
}

@article{Surowka2020,
    title={On duality between Cosserat elasticity and fractons},
    volume={8},
    ISSN={2542-4653},
    url={https://scipost.org/10.21468/SciPostPhys.8.4.065},
    DOI={10.21468/SciPostPhys.8.4.065},
    number={4},
    pages = "065",
    journal={SciPost Physics},
    publisher={SciPost Foundation},
    author={Gromov, Andrey and Sur\'{o}wka, Piotr},
    year={2020},
    month={april}
}

@article{Surowka2021,
  title = {Dual gauge theory formulation of planar quasicrystal elasticity and fractons},
  author = {Sur\'owka, Piotr},
  journal = {Phys. Rev. B},
  volume = {103},
  issue = {20},
  pages = {L201119},
  numpages = {5},
  year = {2021},
  month = {May},
  publisher = {American Physical Society},
  doi = {10.1103/PhysRevB.103.L201119},
  url = {https://link.aps.org/doi/10.1103/PhysRevB.103.L201119}
}

@article{Gaa_2021,
  title = {Fracton-elasticity duality in twisted moir\'e superlattices},
  author = {Gaa, Jonas and Palle, Grgur and Fernandes, Rafael M. and Schmalian, J\"org},
  journal = {Phys. Rev. B},
  volume = {104},
  issue = {6},
  pages = {064109},
  numpages = {9},
  year = {2021},
  month = {Aug},
  publisher = {American Physical Society},
  doi = {10.1103/PhysRevB.104.064109},
  url = {https://link.aps.org/doi/10.1103/PhysRevB.104.064109}
}

@article{Surowka2024a,
  title = {Fracton-elasticity duality on curved manifolds},
  author = {Tsaloukidis, Lazaros and Fern\'andez-Melgarejo, Jos\'e J. and Molina-Vilaplana, Javier and Sur\'owka, Piotr},
  journal = {Phys. Rev. B},
  volume = {109},
  issue = {8},
  pages = {085427},
  numpages = {10},
  year = {2024},
  month = {Feb},
  publisher = {American Physical Society},
  doi = {10.1103/PhysRevB.109.085427},
  url = {https://link.aps.org/doi/10.1103/PhysRevB.109.085427}
}

@article{Surowka2024b,
  title = {Elastic Li\'enard-Wiechert potentials of dynamical dislocations from tensor gauge theory in $2+1$ dimensions},
  author = {Tsaloukidis, Lazaros and Sur\'owka, Piotr},
  journal = {Phys. Rev. B},
  volume = {109},
  issue = {10},
  pages = {104118},
  numpages = {10},
  year = {2024},
  month = {Mar},
  publisher = {American Physical Society},
  doi = {10.1103/PhysRevB.109.104118},
  url = {https://link.aps.org/doi/10.1103/PhysRevB.109.104118}
}

@article{Gromov2019,
  title = {Towards Classification of Fracton Phases: The Multipole Algebra},
  author = {Gromov, Andrey},
  journal = {Phys. Rev. X},
  volume = {9},
  issue = {3},
  pages = {031035},
  numpages = {19},
  year = {2019},
  month = {Aug},
  publisher = {American Physical Society},
  doi = {10.1103/PhysRevX.9.031035},
  url = {https://link.aps.org/doi/10.1103/PhysRevX.9.031035}
}

@article{Nguyen_2020,
   title={Fracton-elasticity duality of two-dimensional superfluid vortex  crystals: defect interactions and quantum melting},
   volume={9},
   ISSN={2542-4653},
   url={http://dx.doi.org/10.21468/SciPostPhys.9.5.076},
   DOI={10.21468/scipostphys.9.5.076},
   number={5},
   pages={076},
   journal={SciPost Physics},
   publisher={Stichting SciPost},
   author={Nguyen, Dung and Gromov, Andrey and Moroz, Sergej},
   year={2020},
   month=nov }

@article{Nguyen_2024b,
    author = "Nguyen, Dung Xuan and Moroz, Sergej",
    title = "{On quantum melting of superfluid vortex crystals: from Lifshitz scalar to dual gravity}",
 
    archivePrefix = "arXiv",
    primaryClass = "cond-mat.quant-gas",
    doi = "10.21468/SciPostPhys.17.6.164",
    journal = "SciPost Phys.",
    volume = "17",
    pages = "164",
    year = "2024"
}

@article{Radzihovsky_2024,
  title = {Quantum vortex lattice: Lifshitz duality, topological defects, and multipole symmetries},
  author = {Du, Yi-Hsien and Lam, Ho Tat and Radzihovsky, Leo},
  journal = {Phys. Rev. B},
  volume = {110},
  issue = {3},
  pages = {035164},
  numpages = {16},
  year = {2024},
  month = {Jul},
  publisher = {American Physical Society},
  doi = {10.1103/PhysRevB.110.035164},
  url = {https://link.aps.org/doi/10.1103/PhysRevB.110.035164}
}

@article{Caddeo_2022,
   title={Emergent dipole gauge fields and fractons},
   volume={106},
   ISSN={2470-0029},
   url={http://dx.doi.org/10.1103/PhysRevD.106.L111903},
   DOI={10.1103/physrevd.106.l111903},
   number={11},
   pages = {L111903},
   journal={Physical Review D},
   publisher={American Physical Society (APS)},
   author={Caddeo, Alessio and Hoyos, Carlos and Musso, Daniele},
   year={2022},
   month=dec }

@article{Pena2023,
  title = {Fractons, symmetric gauge fields and geometry},
  author = {Pe\~na-Ben\'{\i}tez, Francisco},
  journal = {Phys. Rev. Res.},
  volume = {5},
  issue = {1},
  pages = {013101},
  numpages = {7},
  year = {2023},
  month = {Feb},
  publisher = {American Physical Society},
  doi = {10.1103/PhysRevResearch.5.013101},
  url = {https://link.aps.org/doi/10.1103/PhysRevResearch.5.013101}
}

@article{Hirono_2024,
   title={A symmetry principle for gauge theories with fractons},
   volume={16},
   ISSN={2542-4653},
   url={http://dx.doi.org/10.21468/SciPostPhys.16.2.050},
   DOI={10.21468/scipostphys.16.2.050},
   number={2},
   pages={050},
   journal={SciPost Physics},
   publisher={Stichting SciPost},
   author={Hirono, Yuji and You, Minyoung and Angus, Stephen and Cho, Gil Young},
   year={2024},
   month=feb }

@article{Du2024,
  title = {Nonlinear Lifshitz photon theory in condensed matter systems},
  author = {Du, Yi-Hsien and Xu, Cenke and Son, Dam Thanh},
  journal = {Phys. Rev. B},
  volume = {109},
  issue = {3},
  pages = {035135},
  numpages = {9},
  year = {2024},
  month = {Jan},
  publisher = {American Physical Society},
  doi = {10.1103/PhysRevB.109.035135},
  url = {https://link.aps.org/doi/10.1103/PhysRevB.109.035135}
}

@article{glodkowski2025,
  title = {Quadrupole gauge theory: Anti-Higgs mechanism and elastic dual},
  author = {G\l{}\'odkowski, Aleksander and Matus, Pawe\l{} and Pe\~na-Ben\'{\i}tez, Francisco and Tsaloukidis, Lazaros},
  journal = {Phys. Rev. D},
  volume = {112},
  issue = {12},
  pages = {L121702},
  numpages = {6},
  year = {2025},
  month = {Dec},
  publisher = {American Physical Society},
  doi = {10.1103/wnvt-gmwv},
  url = {https://link.aps.org/doi/10.1103/wnvt-gmwv}
}

@article{Low2002,
  title = {Spontaneously Broken Spacetime Symmetries and Goldstone's Theorem},
  author = {Low, Ian and Manohar, Aneesh V.},
  journal = {Phys. Rev. Lett.},
  volume = {88},
  issue = {10},
  pages = {101602},
  numpages = {4},
  year = {2002},
  month = {Feb},
  publisher = {American Physical Society},
  doi = {10.1103/PhysRevLett.88.101602},
  url = {https://link.aps.org/doi/10.1103/PhysRevLett.88.101602}
}

@article{Landry_2020,
   title={The coset construction for non-equilibrium systems},
   volume={2020},
   ISSN={1029-8479},
   url={http://dx.doi.org/10.1007/JHEP07(2020)200},
   DOI={10.1007/jhep07(2020)200},
   number={7},
   pages={200},
   journal={Journal of High Energy Physics},
   publisher={Springer Science and Business Media LLC},
   author={Landry, Michael J.},
   year={2020},
   month=jul }

@misc{landry2022higherformnonstuckelbergsymmetriesnonequilibrium,
      title={Higher-form and (non-)St\"uckelberg symmetries in non-equilibrium systems}, 
      author={Michael J. Landry},
      year={2022},
      eprint={2101.02210},
      archivePrefix={arXiv},
      primaryClass={hep-th},
      url={https://arxiv.org/abs/2101.02210}, 
}

@article{Volkov1973,
title = {Phenomenological lagrangians},
author = {Volkov, D V},
doi = {},
url = {https://www.osti.gov/biblio/4340109}, 
journal = {Sov. J. Particles Nucl.},
number = 1,
pages={1-17},
volume = 4,
place = {United States},
year = {1973},
month = {7}
}

@inproceedings{Ogievetsky1974,
  author    = {Ogievetsky, V. I.},
  title     = {Nonlinear realizations of internal and space-time symmetries},
  year      = {1974},
  booktitle = {Proceedings of the Tenth Winter School
of Theoretical Physics in Karpacz, Vol. 1}
}

@article{Hirono:2021lmd,
    author = "Hirono, Yuji and Qi, Yong-Hui",
    title = "{Effective field theories for gapless phases with fractons via a coset construction}",
    archivePrefix = "arXiv",
    primaryClass = "cond-mat.str-el",
    doi = "10.1103/PhysRevB.105.205109",
    journal = "Phys. Rev. B",
    volume = "105",
    number = "20",
    pages = "205109",
    year = "2022"
}

@article{Nicolis:2013lma,
    author = "Nicolis, Alberto and Penco, Riccardo and Rosen, Rachel A.",
    title = "{Relativistic Fluids, Superfluids, Solids and Supersolids from a Coset Construction}",
    archivePrefix = "arXiv",
    primaryClass = "hep-th",
    doi = "10.1103/PhysRevD.89.045002",
    journal = "Phys. Rev. D",
    volume = "89",
    number = "4",
    pages = "045002",
    year = "2014"
}

@ARTICLE{Popov_1973,
       author = {{Popov}, V.~N.},
        title = "{Quantum vortices and phase transitions in Bose systems}",
      journal = {Soviet Journal of Experimental and Theoretical Physics},
         year = 1973,
        month = aug,
       volume = {37},
        pages = {341},
        url = {http://jetp.ras.ru/cgi-bin/e/index/e/37/2/p341?a=list}
}

@article{Son2005,
  title = {Effective Lagrangian and Topological Interactions in Supersolids},
  author = {Son, D. T.},
  journal = {Phys. Rev. Lett.},
  volume = {94},
  issue = {17},
  pages = {175301},
  numpages = {4},
  year = {2005},
  month = {May},
  publisher = {American Physical Society},
  doi = {10.1103/PhysRevLett.94.175301},
  url = {https://link.aps.org/doi/10.1103/PhysRevLett.94.175301}
}

@article{Watanabe_2013,
  title = {Redundancies in Nambu-Goldstone Bosons},
  author = {Watanabe, Haruki and Murayama, Hitoshi},
  journal = {Phys. Rev. Lett.},
  volume = {110},
  issue = {18},
  pages = {181601},
  numpages = {5},
  year = {2013},
  month = {May},
  publisher = {American Physical Society},
  doi = {10.1103/PhysRevLett.110.181601},
  url = {https://link.aps.org/doi/10.1103/PhysRevLett.110.181601}
}

@article{Kivelson2024,
  title = {Dynamical defects in a two-dimensional Wigner crystal: Self-doping and kinetic magnetism},
  author = {Kim, Kyung-Su and Esterlis, Ilya and Murthy, Chaitanya and Kivelson, Steven A.},
  journal = {Phys. Rev. B},
  volume = {109},
  issue = {23},
  pages = {235130},
  numpages = {13},
  year = {2024},
  month = {Jun},
  publisher = {American Physical Society},
  doi = {10.1103/PhysRevB.109.235130},
  url = {https://link.aps.org/doi/10.1103/PhysRevB.109.235130}
}

@article{Grosvenor:2021hkn,
    author = "Grosvenor, Kevin T. and Hoyos, Carlos and Pe\~na-Benitez, Francisco and Sur\'owka, Piotr",
    title = "{Space-Dependent Symmetries and Fractons}",
    archivePrefix = "arXiv",
    primaryClass = "hep-th",
    doi = "10.3389/fphy.2021.792621",
    journal = "Front. in Phys.",
    volume = "9",
    pages = "792621",
    year = "2022"
}

@article{Nandkishore:2018sel,
    author = "Nandkishore, Rahul M. and Hermele, Michael",
    title = "{Fractons}",
    archivePrefix = "arXiv",
    primaryClass = "cond-mat.str-el",
    doi = "10.1146/annurev-conmatphys-031218-013604",
    journal = "Ann. Rev. Condensed Matter Phys.",
    volume = "10",
    pages = "295--313",
    year = "2019"
}

@article{Pretko:2020cko,
    author = "Pretko, Michael and Chen, Xie and You, Yizhi",
    title = "{Fracton Phases of Matter}",
    archivePrefix = "arXiv",
    primaryClass = "cond-mat.str-el",
    doi = "10.1142/S0217751X20300033",
    journal = "Int. J. Mod. Phys. A",
    volume = "35",
    number = "06",
    pages = "2030003",
    year = "2020"
}

@article{Gromov:2022cxa,
    author = "Gromov, Andrey and Radzihovsky, Leo",
    title = "{Colloquium: Fracton matter}",
    archivePrefix = "arXiv",
    primaryClass = "cond-mat.str-el",
    doi = "10.1103/RevModPhys.96.011001",
    journal = "Rev. Mod. Phys.",
    volume = "96",
    number = "1",
    pages = "011001",
    year = "2024"
}

@article{Pretko2018c,
  title = {Symmetry-Enriched Fracton Phases from Supersolid Duality},
  author = {Pretko, Michael and Radzihovsky, Leo},
  journal = {Phys. Rev. Lett.},
  volume = {121},
  issue = {23},
  pages = {235301},
  numpages = {7},
  year = {2018},
  month = {Dec},
  publisher = {American Physical Society},
  doi = {10.1103/PhysRevLett.121.235301},
  url = {https://link.aps.org/doi/10.1103/PhysRevLett.121.235301}
}

@article{Jain:2021ibh,
    author = "Jain, Akash and Jensen, Kristan",
    title = "{Fractons in curved space}",
    archivePrefix = "arXiv",
    primaryClass = "hep-th",
    doi = "10.21468/SciPostPhys.12.4.142",
    journal = "SciPost Phys.",
    volume = "12",
    number = "4",
    pages = "142",
    year = "2022"
}

@article{Kumar2019,
  title = {Symmetry-enforced fractonicity and two-dimensional quantum crystal melting},
  author = {Kumar, Ajesh and Potter, Andrew C.},
  journal = {Phys. Rev. B},
  volume = {100},
  issue = {4},
  pages = {045119},
  numpages = {13},
  year = {2019},
  month = {Jul},
  publisher = {American Physical Society},
  doi = {10.1103/PhysRevB.100.045119},
  url = {https://link.aps.org/doi/10.1103/PhysRevB.100.045119}
}

\newpage
\phantom{a}
\newpage
\setcounter{figure}{0}
\setcounter{equation}{0}
\setcounter{section}{0}

\renewcommand{\thetable}{S\arabic{table}}
\renewcommand{\thefigure}{S\arabic{figure}}
\renewcommand{\theequation}{S\arabic{equation}}
\renewcommand{\thepage}{S\arabic{page}}

\renewcommand{\thesection}{S\arabic{section}}

\onecolumngrid

\begin{center}
{\large \bf Supplemental Material:\\
Fracton-elasticity duality in Wigner crystals}\\

\vspace{0.3cm}

\setcounter{page}{1}
\end{center}

\label{pagesupp}

\section{Coset construction}
\label{app:coset}

In order to construct the effective action, we make use of the coset construction \cite{Volkov1973,Ogievetsky1974,Low2002,Landry_2020,Hirono:2021lmd}. The preliminary step is determining the commutation relations between the generators of the symmetries defined in Sec.~\ref{sec:symmetries}. Using Eqs.~(\ref{eq:magnetic_symmetry}), (\ref{eq:Ti_symmetry}) and~(\ref{eq:Pi_symmetry}), we find the following generator algebra:
\begin{equation}
    \begin{split}
        [P_i,P_j]&=[T_i,T_j]=0,\\ [T_i,P_j]&=\frac{n_0}{2}\epsilon_{ijk}dx_k \mathcal{Q}^{(1)}.
        \end{split}
\end{equation}
The second commutator is understood in the following way: an application of the symmetry operation $e^{-\xi^T_iT_i}e^{-\xi^P_iP_i}e^{\xi^T_iT_i}e^{\xi^P_iP_i}$ with infinitesimal $\xi^T_i$, $\xi^P_i$ is equivalent to applying $\exp\left(\xi^T_i\xi^P_j\frac{n_0}{2}\epsilon_{ijk}dx_k \mathcal{Q}^{(1)}\right)$ defined in Eq.~(\ref{eq:magnetic_symmetry}). The noncommutativity of the two types of translations, $T_i$ and $P_i$, will play a crucial role in constructing the effective action of the Wigner crystal\footnote{In Ref. \cite{glodkowski2025} it was assumed that $[T_i,T_j]\neq 0$. However, the operators $T_i$ employed in this paper are linear combinations of the operators $\tilde{P}_i$ and $T_i$ defined in Ref. \cite{glodkowski2025}.}. Furthermore, the operators $H$ and $\mathcal{Q}^{(1)}$ are found to commute with all the generators. Finally, regarding the rotation generators $J_i$ and $L_i$, their commutators turn out to play no role in the construction of the effective action, and consequently we do not display them here. 

To define the coset element, we first introduce the Stueckelberg fields $u_i(x)$ and $\phi(x)=\phi_\mu(x) dx_\mu$ corresponding to the broken generators $T_i$ and $\mathcal{Q}^{(1)}$, respectively. Following the discussion in Sec.~\ref{sec:effective}, we do not define a separate field for the broken rotations. The coset element in the presence of both 1-form and 0-form symmetries is defined for a one-dimensional curve $\mathcal{C}$ and a point $x$ as \cite{landry2022higherformnonstuckelbergsymmetriesnonequilibrium}
\begin{equation}
    g(\mathcal{C},x) = e^{\mathcal{Q}^{(1)}\int_\mathcal{C} \phi}e^{tH}e^{x_iP_i}e^{u_i(x)T_i}.
    \label{eq:coset}
\end{equation}
The symmetry group acts on the coset element (\ref{eq:coset}) by left multiplication. We find
\begin{subequations}
\begin{align}
    \exp(\omega\mathcal{O}^{(1)}):&\qquad \phi \rightarrow \phi+\omega,\\
    \exp(\xi_iT_i):&\qquad \phi\rightarrow\phi+\frac{n_0}{2}\epsilon_{ijk}\xi_ix_jdx_k,\quad u_i \rightarrow u_i+\xi_i,\\
    \exp(\xi_iP_i):&\qquad x_i\rightarrow x_i+\xi_i, \\
    \exp(\xi H):&\qquad t\rightarrow t+\xi.
\end{align}
\end{subequations}
Thus, the fields transform in a way consistent with their physical interpretation given in Sec.~\ref{sec:effective}. 

\begin{figure}[h!]
    \includegraphics[width=0.25\linewidth]{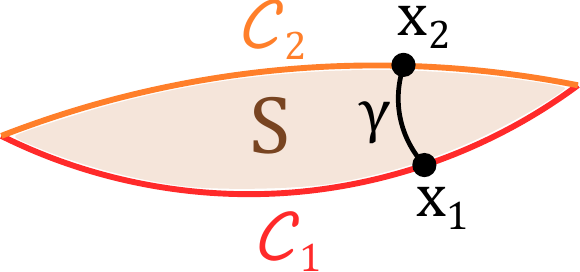}
    \caption{Points and curves used in the calculation of the Maurer-Cartan form.}
    \label{fig:contours}
\end{figure}

Finally, invariant combinations of the fields can be found by calculating the Maurer-Cartan form in the following way \cite{landry2022higherformnonstuckelbergsymmetriesnonequilibrium}. We take two infinitesimally close lines $\mathcal{C}_1$ and $\mathcal{C}_2$ such that the curve $\mathcal{C}_2\cup (-\mathcal{C}_1)$, with minus sign indicating reversing the orientation, is the boundary of some surface $S$: $\partial S=\mathcal{C}_2\cup (-\mathcal{C}_1)$. Similarly, we take two points $x_1$ and $x_2$ such that for some path $\gamma$ we have $\partial\gamma=x_2\cup (-x_1)$, see Fig.~\ref{fig:contours}. We then find the 2-form $\Omega_{\mathcal{Q}^{(1)}}$ and 1-forms $\Omega_{T_i}$ and $\Omega_{\theta_i}$ such that
\begin{equation}
    g^{-1}(\mathcal{C}_1,x_1)g(\mathcal{C}_2,x_2)
    =e^{\mathcal{Q}^{(1)}\int_S \Omega_{\mathcal{Q}^{(1)}}}e^{H\int_\gamma dt}e^{P_i\int_\gamma dx_i}e^{T_i\int_\gamma \Omega_{T_i}}.
    \label{eq:MC_definition}
\end{equation}
The forms $\Omega_{\mathcal{Q}^{(1)}}$ and $\Omega_{T_i}$ are invariant under all the symmetries by construction, and can be used as building blocks of the effective action. The left hand side of Eq.~(\ref{eq:MC_definition}) can be calculated explicitly using the Baker-Campbell-Hausdorff formula, and we find
\begin{equation}
\begin{split}
    \Omega_{\mathcal{Q}^{(1)}} &= 2\partial_{[t} \phi_{i]}dt\wedge dx_i + \left(2\partial_{[i}\phi_{j]}-n_0\epsilon_{ijk}u_k\right)dx_i\wedge dx_j,\\
    \Omega_{T_i} & = du_i.
\end{split}
\end{equation}
The different components of the forms above are precisely the objects in Eq.~(\ref{eq:invariants}).

\section{Derivation of the duality}
\label{app:duality}

We start by writing down the Hubbard-Stratonovich transformation of the action~(\ref{eq:action}):
\begin{equation}
\begin{split}
    S_{\mathrm{HS}} =  \frac12 
    \int d^4 x &\left[-c_B^{-1} K_{it}^2 + \frac{c_E^{-1}}{2}K_{ij}^2-c_{u}^{-1}\tau_{it}^2 +\tilde{C}_{ijkl}\tau_{ij}\tau_{kl}-2K_{it}\left(2\partial_{[t}\phi_{i]}\right)\right.\\
    &\left.-K_{ji}\left(2\partial_{[i} \phi_{j]}-n_0\epsilon_{ijk}u_k\right)-2\tau_{it}\left(\partial_tu_i\right)-2\tau_{ji}\partial_i u_{j}+\epsilon_{ijk}\tau_{jk}\zeta_i\right],
    \label{eqapp:action_HS}
\end{split}
\end{equation}
where $K_{it}$, $K_{ij}$, $\tau_{it}$, $\tau_{ij}$ are Hubbard-Stratonovich fields, the inverse elasticity tensor is
\begin{equation}
    \tilde{C}_{ijkl} = \frac{C_b^{-1}}{3} \delta_{ij}\delta_{kl}+\frac{C_s^{-1}}{2}(\delta_{ik}\delta_{jl}+\delta_{il}\delta_{jk}-\frac{2}{3}\delta_{ij}\delta_{kl}),
\end{equation}
and we have added a set of Lagrange multipliers $\zeta_i$ which constrain the stress tensor to be symmetric: $\epsilon_{ijk}\tau_{jk}=0.$ Now we split the fields $\phi$ and $u_i$ into smooth parts denoted with a bar and singular parts, related to lattice defects, denoted with the superscript $s$:
\begin{equation}
    \phi_\mu = \bar\phi_\mu+\phi^s_\mu, \qquad u_i=\bar{u}_i+u_i^s.
\end{equation}
$\zeta_i$ are taken to be smooth, $\zeta_i=\bar \zeta_i$. Integrating out the smooth fields in the action~(\ref{eqapp:action_HS}) imposes a set of constraints, which are most conveniently written using spacetime indices $\mu$:
\begin{equation}
\begin{split}
    \partial_\nu K_{\mu\nu}&=0,\\
    \partial_\mu\tau_{i\mu}&=\frac12 n_0\epsilon_{ijk}K_{jk},\\
    0&=\epsilon_{ijk}\tau_{jk}.
\end{split}\label{eqapp:constraints}
\end{equation}
Defining the differential forms $K=\frac12 K_{\mu\nu}dx_\mu\wedge dx_\nu$, $\tau_i=\tau_{i\mu}dx_\mu$, the constraints in Eq.~(\ref{eqapp:constraints}) can be conveniently expressed as
\begin{equation}
\begin{split}
    d*K&=0,\\
    d*\tau_i &= \frac{1}{2}n_0\epsilon_{ijk}dx_j\wedge dx_k\wedge * K,\\
    0&=\epsilon_{ijk}dx_j\wedge*\tau_{k}.
\end{split}\label{eqapp:constraints2}
\end{equation}
The most general solution to Eq.~(\ref{eqapp:constraints2}) has the form
\begin{equation}
\begin{split}
    K&=*\left(dA^{(\phi)}\right),\\
    \tau_i &= *\left(dA^{(u)}_i-\frac{1}{2}n_0\epsilon_{ijk}dx_j\wedge dx_k\wedge A^{(\phi)}\right),
\end{split} \label{eqapp:field_strengths}
\end{equation}
for some 1-form $A^{(\phi)}\equiv A^{(\phi)}_\mu dx_\mu$ and a 2-form $A^{(u)}_i\equiv \frac12 A^{(u)}_{i\mu\nu} dx_\mu\wedge dx_\nu$, subject to the constraint
\begin{equation}
\begin{split}
    \epsilon_{ijk}dx_j\wedge A_k^{(u)}=-\frac{1}{4}d A^{(\Theta)}_i
\end{split} \label{eqapp:field_strengths2}
\end{equation}
for some 2-form field $A^{(\Theta)}_i\equiv \frac12 A^{(\Theta)}_{i\mu\nu}dx_\mu \wedge dx_\nu$. The equalities in Eqs.~(\ref{eqapp:field_strengths}) and~(\ref{eqapp:field_strengths2}) do not determine $A^{(\phi)}$, $A^{(u)}_i$ and $A^{(\Theta)}_i$ unambiguously; rather, there is a gauge redundancy
\begin{equation}
\begin{split}
    A^{(\phi)}&\rightarrow A^{(\phi)}+d\lambda^{(\phi)},\\
    A^{(u)}_i&\rightarrow A^{(u)}_i+d\lambda^{(u)}_i+\frac12 n_0\epsilon_{ijk}dx_j\wedge dx_k \lambda^{(\phi)},\\
    A^{(\Theta)}_i&\rightarrow A^{(\Theta)}_i+d\lambda^{(\Theta)}_i+4\epsilon_{ijk}dx_j\wedge \lambda^{(u)}_{k}
\end{split} \label{eqapp:gauge}
\end{equation}
for an arbitrary 0-form (scalar) $\lambda^{(\phi)}$ and 1-forms $\lambda^{(u)}_i$ and $\lambda^{(\Theta)}_i$. 

The Hubbard-Stratonovich action (\ref{eqapp:action_HS}) can be split into the electromagnetic part, which describes the physics of the smooth gauge fields, and the part coupling the gauge fields to the defect currents:
\begin{equation}
    S_{\mathrm{dual}}=S_{\mathrm{EM}}+S_{\mathrm{coupling}}
\end{equation}
with
\begin{equation}
    S_{\mathrm{EM}}=  \frac12 
    \int d^4 x \left[-c_B^{-1} K_{it}^2 + \frac{c_E^{-1}}{2}K_{ij}^2-c_{u}^{-1}\tau_{it}^2 +\tilde{C}_{ijkl}\tau_{ij}\tau_{kl}\right]
    \label{eqapp:SEM}
\end{equation}
and
\begin{equation}
\begin{split}
    S_{\mathrm{coupling}} =   
    \int d^4 x &\left[\left(d\phi^{s}-\frac12 n_0\epsilon_{ijk}u_k^{s}dx_i\wedge dx_j \right)\wedge *K-du_i^{s}\wedge* \tau_i\right].
\end{split} \label{eqapp:action_coupling_t}
\end{equation}
Expressing the field strengths by their formulas in Eq.~(\ref{eqapp:field_strengths}) and applying integration by parts gives
\begin{equation}
\begin{split}
    S_{\mathrm{coupling}} =   
    \int d^4 x \left[A^{(\phi)}\wedge *J^{\mathrm{vac}}+A^{(u)}_i\wedge *J^{\mathrm{disl}}_i\right],
\end{split} \label{eqapp:action_coupling}
\end{equation}
where the two defect currents $J^{\mathrm{vac}}=J^{\mathrm{vac}}_\mu dx_\mu$ and the dislocation current $J^{\mathrm{disl}}_i=\frac12 J^{\mathrm{disl}}_{i\mu\nu}dx_\mu\wedge dx_\nu$ are defined as
\begin{equation}
\begin{split}
    J^{\mathrm{vac}}_\mu & = \epsilon_{\mu\nu\gamma\delta}\partial_\nu \partial_\gamma \phi^s_\delta\,,\\
    J^{\mathrm{disl}}_{i\mu\nu} & = \epsilon_{\mu\nu\gamma\delta}\partial_\gamma \partial_\delta u^s_i\,. \label{eqapp:defects}
\end{split}
\end{equation}
The symmetry of the action under the gauge transformation~(\ref{eqapp:gauge}) then enforces the continuity equations
\begin{align}
    d * J^{\mathrm{disl}}_i&=0,\label{eqapp:cont_disl}\\
    d * J^{\mathrm{vac}}-n_0\epsilon_{ijk}dx_j \wedge dx_k\wedge *J^{\mathrm{disl}}_{i}&=0.\label{eqapp:cont_vac}
\end{align}
The continuity equations~(\ref{eq:cont1}) and (\ref{eq:cont2}) are the different components of Eq.~(\ref{eqapp:cont_disl}), while Eq.~(\ref{eq:cont3}) can be obtained by combining~(\ref{eqapp:cont_vac}) with~(\ref{eq:cont2}).

Simultaneously, an explicit evaluation of Eq.~(\ref{eqapp:field_strengths2}) gives
\begin{equation}
\begin{split}
    A^{(u)}_{ijt} & = -\frac{1}{4}\epsilon_{jkl}\epsilon_{lmn}\epsilon_{ikp}\left(\partial_{t}A^{(\Theta)}_{pmn}+2\partial_mA^{(\Theta)}_{pnt}\right), \\
    A^{(u)}_{ikl} & = \tilde{A}^{(u)}_{(ij)}\epsilon_{jkl}+\frac{1}{8}\epsilon_{jkl}\epsilon_{lmn}\epsilon_{ijp}\partial_{l}A^{(\Theta)}_{pmn},
\end{split} \label{eqapp:ATheta}
\end{equation}
where $\tilde{A}^{(u)}_{(ij)}$ is a symmetric tensor independent of $A^{(\Theta)}_i$. Plugging this result into Eq.~(\ref{eqapp:action_coupling}) and integrating by parts, we can isolate a coupling of the form $A^{(\Theta)}_i\wedge*J_i^{\mathrm{disc}}$ with 
\begin{equation}
    J^{\mathrm{disc}}_{i\mu\nu} = \frac{1}{2}\epsilon_{ijk}\partial_j J^{\mathrm{disl}}_{k\mu\nu} = \frac{1}{2}\epsilon_{ijk}\epsilon_{\mu\nu\gamma\delta}\partial_\gamma \partial_\delta \partial_j u_k^{s}=\epsilon_{\mu\nu\gamma\delta}\partial_\gamma \partial_\delta \Theta_i^s,
\end{equation}
where the bond angle vector $\Theta_i$ is defined as $\Theta_i = \frac12 \epsilon_{ijk}\partial_ju_k^s$, see Eq.~(\ref{eq:bond_angle}). The symmetry under the gauge transformation $\lambda^{(\Theta)}$ in Eq.~(\ref{eqapp:gauge}) then gives rise to an additional continuity equation,
\begin{equation}
    d*J_i^{\mathrm{disc}}=0.
\end{equation}

\section{Anti-Higgs mechanism}
\label{app:higgs}

In this section we will find the gapless modes of the action~(\ref{eq:action_Higgs}) first in the absence of, and then in the presence of, the vacancy condensate. While the most straightforward way to do it is by solving the full set of equations of motion, we choose to first integrate out the gapped degrees of freedom in order to gain insight into the structure of the low-energy effective action.

In the symmetric phase ($\rho_s=0$) we first focus on the following terms in (\ref{eq:SEM}):
\begin{equation}
    -\frac{1}{2}c_{u}^{-1}\tau_{it}^2 +\frac{1}{6}C_b^{-1}\tau_{ii}^2 = -\frac{c_{u}^{-1}}{2}\left(n_0A^{(\phi)}_i-3\partial_{[x}A^{(u)}_{i;yz]}\right)^2+\frac{C_b^{-1}}{6}\left(3n_0A_t^{(\phi)}-\frac{3}{2}\epsilon_{ijk}\partial_{[t}A^{(u)}_{i;jk]}\right)^2.
\end{equation}
Since both of these terms are massive, at low enough frequencies they can be integrated out of the action, fixing the components of $A^{(\phi)}$ in terms of derivatives of $A^{(u)}_i$. Then, at long wavelengths the term $\frac{c_E^{-1}}{4}K_{ij}K_{ij}$ is subleading with respect to $\frac{C_s^{-1}}{4}\tau_{\langle ij\rangle}\tau_{\langle ij\rangle}$. For ease of notation, let us define
\begin{equation}
    F_{ij} \equiv \epsilon_{ijk}K_{kt},\qquad \tilde{A}_{ij} \equiv \frac12\epsilon_{jkl}A^{(u)}_{ikl}.
\end{equation}
Keeping only the leading-order terms, we find that the low-energy effective action reads
\begin{equation}
    S_{\mathrm{low-energy}}=\frac{1}{2}\int d^4x \left[-\frac{c_B^{-1}}{2}F_{ij}F_{ij}+C_s^{-1}\tau_{\langle ij \rangle}\tau_{\langle ij\rangle}\right],
    \label{eqapp:S_low}
\end{equation}
where 
\begin{equation}
\begin{split}
    F_{ij} & = n_0^{-1}\left(\partial_i\partial_k\tilde{A}_{jk}-\partial_j\partial_k\tilde{A}_{ik}\right),\\
    \tau_{\langle ij\rangle} &= \partial_t\tilde{A}_{\langle ij\rangle}+\partial_k\left(\epsilon_{kl\langle i}A^{(u)}_{j\rangle lt}\right).
\end{split}
\end{equation}
Furthermore, because of the equality~(\ref{eqapp:ATheta}), $A^{(u)}_{ijt}$ and $\tilde{A}_{[ij]}$ are equal to various combinations of derivatives $A^{(\Theta)}_{i\mu\nu}$, meaning that the independent fields in Eq.~(\ref{eqapp:S_low}) are actually $\tilde{A}_{(ij)}$ and $A^{(\Theta)}_{i\mu\nu}$. Note, however, that the trace of $\tilde{A}_{ij}$ is absent from the action~(\ref{eqapp:S_low}); while this might seem surprising at first, it is a consequence of the fact that the only fields that vary under the gauge parameter $\lambda^{(\phi)}$ are $A^{(\phi)}_\mu$ and $\tilde{A}_{ii}$, so eliminating $A^{(\phi)}_\mu$ from the action necessarily eliminates $\tilde{A}_{ii}$ as well. Now, varying the action~(\ref{eqapp:S_low}) with respect to $\tilde{A}_{\langle ij\rangle}$ gives the ``Amp\`{e}re's law"
\begin{equation}
    \partial_t\tau_{\langle ij\rangle}-\frac{C_s}{4n_0c_B}\partial_k\partial_{\langle i}F_{j\rangle k}=0.
    \label{eqapp:ampere1}
\end{equation}
Varying the action~(\ref{eqapp:S_low}) with respect to $A^{(\Theta)}_{ijt}$ gives the ``Gauss's law"
\begin{equation}
    \epsilon_{ikl}\epsilon_{jmn}\partial_k \partial_m \tau_{\langle nl \rangle}=0.
    \label{eqapp:gauss1}
\end{equation}
Varying the action~(\ref{eqapp:S_low}) with respect to $A^{(\Theta)}_{ijk}$ gives combinations of derivatives of the Amp\`{e}re's law, and therefore provides no new information. Incidentally, this shows that the low-energy gauge theory can be simplified by gauge-fixing $A^{(\Theta)}_{ijk}=0$ before calculating variations of the action. Similarly, since Eq.~(\ref{eqapp:gauss1}) is symmetric in $i$ and $j$, gauge-fixing $A^{(\Theta)}_{[ij]t}=0$ at the level of the action does not change the equations of motion either. Thus, an equivalent but simpler gauge theory can be formulated using only the traceless symmetric tensor $\tilde{A}_{\langle ij \rangle}$ and the symmetric tensor $A^{(\Theta)}_{(ij)t}$ with the definitions
\begin{equation}
\begin{split}
    F_{ij} & = n_0^{-1}\left(\partial_i\partial_k\tilde{A}_{\langle jk\rangle}-\partial_j\partial_k\tilde{A}_{\langle ik\rangle}\right),\\
    \tau_{\langle ij\rangle} &= \partial_t\tilde{A}_{\langle ij\rangle}+\frac12\epsilon_{mn\langle i}\epsilon_{j\rangle kp}\partial_k\partial_m A^{(\Theta)}_{(pn)t}.
\end{split}
\end{equation}
The gauge transformations of the remaining gauge fields are parametrized by the symmetric tensor $\lambda^{(\Theta)}_{(ij)}$ and the vector $\lambda^{(\Theta)}_{i0}$ and read
\begin{equation}
\begin{split}
    \tilde{A}_{\langle ik\rangle} & \rightarrow \tilde{A}_{\langle ik\rangle} + \frac12 \epsilon_{mn\langle i}\epsilon_{j\rangle kp}\partial_k\partial_m \lambda^{(\Theta)}_{(pn)},\\
    A^{(\Theta)}_{(ij)t} & \rightarrow A^{(\Theta)}_{(ij)t} - \partial_t \lambda^{(\Theta)}_{(ij)} + \partial_{(i}\lambda^{(\Theta)}_{j)0}.
\end{split}
\end{equation}

After finding the equations of motion~(\ref{eqapp:ampere1}) and~(\ref{eqapp:gauss1}), we can further gauge-fix $A^{(\Theta)}_{ijt}=0$, such that $F_{ij}$ and $\tau_{\langle ij\rangle}$ are given in terms of derivatives of $\tilde{A}_{\langle ij\rangle}$ only. Furthermore, assuming that the perturbation propagates along the $x$ direction, we get from Eq.~(\ref{eqapp:gauss1}) $\tilde{A}_{\langle yy\rangle}=\tilde{A}_{\langle zz\rangle}=\tilde{A}_{\langle yz\rangle}=0$, leaving $\tilde{A}_{\langle xy\rangle}$ and $\tilde{A}_{\langle xz\rangle}$ as the only propagating degrees of freedom. The dispersion relation of both modes reads
\begin{equation}
    \omega = \pm \sqrt{\frac{C_s}{2n_0^2 c_B}}k^2.
\end{equation}

Now let us study what changes once the vacancy condensate forms ($\rho_s >0$). We first pay attention to the following terms in the action~(\ref{eq:action_Higgs}):
\begin{equation}
\begin{split}
    &-\frac{c_{u}^{-1}n_0^2}{2}\left(A^{(\phi)}_i-3n_0^{-1}\partial_{[x}A^{(u)}_{i;yz]}\right)^2+\frac{3C_b^{-1}n_0^2}{2}\left(A_t^{(\phi)}-\frac{1}{2}n_0^{-1}\epsilon_{ijk}\partial_{[t}A^{(u)}_{i;jk]}\right)^2+\frac{\rho_s}{2c^{*2}}\left(\partial_t\theta-A^{(\phi)}_t\right)^2-\frac{\rho_s}{2}\left(\partial_i\theta-A^{(\phi)}_i\right)^2 \\
    = & -\frac{M_1}{2}\left(A^{(\phi)}_i-\frac{3c_u^{-1}n_0}{M_1}\partial_{[x}A^{(u)}_{i;yz]}-\frac{\rho_s}{M_1}\partial_i\theta\right)^2-\frac{c_u^{-1}n_0^2\rho_s}{2M_1}\left(3n_0^{-1}\partial_{[x}A^{(u)}_{i;yz]}-\partial_i\theta\right)^2 \\
    & +\frac{M_2}{2}\left(A_t^{(\phi)}-\frac{3C_b^{-1}n_0}{2M_2}\epsilon_{ijk}\partial_{[t}A^{(u)}_{i;jk]}-\frac{\rho_s}{M_2 c^{*2}}\partial_t\theta\right)^2+\frac{3C_b^{-1}n_0^2\rho_s}{2c^{*2}M_2}\left(\frac12n_0^{-1}\epsilon_{ijk}\partial_{[t}A^{(u)}_{i;jk]}-\partial_t\theta\right)^2,
    \label{eqapp:longass}
\end{split}
\end{equation}
where 
\begin{equation}
    M_1 \equiv c_u^{-1}n_0^2+\rho_s,\qquad M_2 \equiv 3C_b^{-1}n_0^2+\frac{\rho_s}{c^{*2}}.
\end{equation}
The first and the third term on the RHS of Eq.~(\ref{eqapp:longass}) can be integrated out of the action at low frequencies, leaving the second and the fourth term behind. Let us further denote
\begin{equation}
    K_1 \equiv \frac{c_u^{-1}n_0\rho_s}{M_1},\qquad K_0 \equiv \frac{3C_b^{-1}n_0\rho_s}{c^{*2}M_2},\qquad T_i\equiv 3n_0^{-1}\partial_{[x}A^{(u)}_{i;yz]}-\partial_i\theta,\qquad T_t\equiv \frac12n_0^{-1}\epsilon_{ijk}\partial_{[t}A^{(u)}_{i;jk]}-\partial_t\theta.
\end{equation}
Keeping only the leading-order terms, we find that the low-energy effective action reads
\begin{equation}
    S_{\mathrm{low-energy}}=\frac{1}{2}\int d^4x \left[-n_0K_1T_i^2+n_0K_0T_t^2+C_s^{-1}\tau_{\langle ij \rangle}\tau_{\langle ij\rangle}\right].
    \label{eqapp:S_low2}
\end{equation}
Varying the action~(\ref{eqapp:S_low2}) with respect to $\tilde{A}_{\langle ij\rangle}$ and $\theta$ (or, equivalently, $\tilde{A}_{ii}$) gives the ``Amp\`{e}re's law"
\begin{equation}
\begin{split}
    C_s^{-1}\partial_t\tau_{\langle ij\rangle}-K_1\partial_{\langle i}T_{j\rangle}& =0,\\
    K_0\partial_tT_t-K_1\partial_iT_i&=0.
\end{split}  \label{eqapp:ampere2}
\end{equation}
Varying the action~(\ref{eqapp:S_low2}) with respect to $A^{(\Theta)}_{ij0}$ gives the ``Gauss's law"
\begin{equation}
    \epsilon_{ikl}\epsilon_{jmn}\partial_k \partial_m \left(C_s^{-1}\tau_{\langle nl \rangle}+\frac{K_0}{3}\delta_{nl}T_t\right)=0.
    \label{eqapp:gauss2}
\end{equation}
Varying the action~(\ref{eqapp:S_low2}) with respect to $A^{(\Theta)}_{ijk}$ gives combinations of derivatives of the Amp\`{e}re's law, and therefore provides no new information. To proceed, use the gauge freedom to set $A^{(\Theta)}_i=0$ and $\tilde{A}_{ii}=0$, such that $T_t$, $T_i$ and $\tau_{\langle ij\rangle}$ are given in terms of derivatives of $\tilde{A}_{\langle ij\rangle}$ and $\theta$ only. Furthermore, assuming that the perturbation propagates along the $x$ direction, we get from Eq.~(\ref{eqapp:gauss2}) $\tilde{A}_{\langle yz\rangle}=0$ and $\tilde{A}_{\langle xx\rangle}=-\frac{2C_sK_0}{3}\theta$, leaving $\tilde{A}_{\langle xy\rangle}$, $\tilde{A}_{\langle xz\rangle}$ and $\theta$ as the only independent propagating degrees of freedom. The dispersion relations read
\begin{equation}
\begin{split}
    \mathrm{For~}\theta:\qquad& \omega = \pm \sqrt{\frac{K_1}{K_0}}k=\pm \sqrt{\frac{3 n_0^2 c^{*2}+C_b \rho_s}{3 n_0^2 +c_u \rho_s}}k, \\
    \mathrm{For~}\tilde{A}_{\langle xy\rangle},\tilde{A}_{\langle xz\rangle}:\qquad& \omega = \pm \sqrt{\frac{C_sK_1}{2n_0}}k= \pm \sqrt{\frac{C_s\rho_s}{2\left(n_0^2+c_u\rho_s\right)}}k.
\end{split}
\end{equation}

\end{document}